\pdfoutput=1

\documentclass[11pt,twoside,a4paper,cmspaper,final,collab]{cms-tdr}
\def\svnVersion{304918}\def\cmsCernNo{2013-037}\def\cmsCernDate{\today}\def\cmsMessage{Published in Physical Review C as \href{http://dx.doi.org/10.1103/PhysRevC.92.034911}{\doi{10.1103/PhysRevC.92.034911.}}}

\begin{document}\cmsNoteHeader{HIN-14-012}

\hyphenation{had-ron-i-za-tion}
\hyphenation{cal-or-i-me-ter}
\hyphenation{de-vices}
\RCS$Revision: 267855 $
\RCS$HeadURL: svn+ssh://svn.cern.ch/reps/tdr2/papers/HIN-14-012/trunk/HIN-14-012.tex $
\RCS$Id: HIN-14-012.tex 267855 2014-11-14 22:08:15Z davidlw $
\newlength\cmsFigWidth
\ifthenelse{\boolean{cms@external}}{\setlength\cmsFigWidth{0.9\textwidth}}{\setlength\cmsFigWidth{0.9\textwidth}}
\ifthenelse{\boolean{cms@external}}{\providecommand{\cmsLeft}{top}}{\providecommand{\cmsLeft}{left}}
\ifthenelse{\boolean{cms@external}}{\providecommand{\cmsRight}{bottom}}{\providecommand{\cmsRight}{right}}
\newcommand {\avg}[1]{\ensuremath{\langle\kern-1.0pt\langle#1\rangle\kern-1.0pt\rangle}}
\newcommand{\deta}    {\ensuremath{\Delta\eta}}
\newcommand{\dphi}    {\ensuremath{\Delta\phi}}
\newcommand{\noff}   {\ensuremath{N_\text{trk}^\text{offline}}\xspace}
\newcommand{\pp}   {\mbox{pp}\xspace}
\newcommand{\Pb}   {\mbox{Pb}\xspace}
\newcommand{\PbPb} {\mbox{PbPb}\xspace}
\newcommand{\pPb}  {\mbox{pPb}\xspace}
\newcommand{\pttrg}      {\ensuremath{\pt^{a}}\xspace}
\newcommand{\ptass}      {\ensuremath{\pt^{b}}\xspace}
\newcommand{\etatrg}      {\ensuremath{\eta^{a}}\xspace}
\newcommand{\etaass}      {\ensuremath{\eta^{b}}\xspace}
\newcommand {\rootsNN}  {\ensuremath{\sqrt{s_{_\mathrm{NN}}}}\xspace}
\providecommand{\EPOS} {\textsc{epos}\xspace}

\cmsNoteHeader{HIN-14-012}
\title{Evidence for transverse momentum and pseudorapidity dependent event plane
fluctuations in \PbPb and \pPb collisions}

\date{\today}

\abstract{
A systematic study of the factorization of long-range azimuthal two-particle correlations into a product
of single-particle anisotropies is presented as a function of \pt and $\eta$ of both particles, and as a function
of the particle multiplicity in \PbPb and \pPb collisions. The data were taken with the CMS detector
for \PbPb collisions at
$\rootsNN = 2.76$\TeV
and \pPb collisions at $\rootsNN = 5.02\TeV$,
covering a very wide range of multiplicity.
Factorization is observed to be broken as a function of both particle \pt and $\eta$.
When measured with particles of different \pt,
the magnitude of the factorization breakdown for the second Fourier harmonic
reaches 20\% for very central \PbPb collisions but decreases rapidly as
the multiplicity decreases. The data are consistent with viscous hydrodynamic
predictions, which suggest that the effect of factorization breaking is mainly sensitive to the initial-state
conditions rather than to the transport properties (e.g., shear viscosity) of the medium.
The factorization breakdown is also computed with particles of different $\eta$.
The effect is found to be
weakest for mid-central \PbPb events but becomes larger for more central or peripheral \PbPb collisions,
and also for very high-multiplicity \pPb collisions. The $\eta$-dependent factorization data provide
new insights to the longitudinal evolution of the medium formed in heavy ion collisions.
}

\hypersetup{%
pdfauthor={CMS Collaboration},%
pdftitle={Evidence for transverse momentum and pseudorapidity dependent event plane
fluctuations in PbPb and pPb collisions},%
pdfsubject={CMS},%
pdfkeywords={CMS, heavy ion, two-particle correlation, factorization,
flow}}

\maketitle
\section{Introduction}

The goal of experiments with heavy ion collisions at ultra-relativistic
energies is to study nuclear matter under extreme conditions.
By studying the azimuthal anisotropy of emitted particles in such collisions,
experiments at the Relativistic Heavy Ion Collider at BNL (RHIC) indicated 
that
a strongly-coupled hot and dense medium is created, which
exhibits a strong collective flow behavior~\cite{BRAHMS,PHOBOS,STAR,PHENIX}.
At the significantly higher collision
energies achieved at the Large Hadron Collider (LHC),
the collective phenomena of this quark gluon plasma have also been studied
in great
detail~\cite{Aamodt:2011by,Abelev:2014pua,ATLAS:2012at,Aad:2013xma,Aad:2014fla,Chatrchyan:2012wg,Chatrchyan:2012ta,Chatrchyan:2013kba,CMS:2013bza}.

The collective expansion of the hot medium in heavy ion collisions can be described by hydrodynamic
flow models. Motivated by such models, the azimuthal distribution of
emitted particles can be characterized by the Fourier components
of the hadron yield distribution in azimuthal angle ($\phi$)~\cite{Ollitrault:1993ba,Voloshin:1994mz,Poskanzer:1998yz},
\begin{linenomath}
\begin{equation}
\label{eq:fourier_single}
\frac{\rd{}N}{\rd\phi} \propto 1+2 \sum_{n}^{} v_{n}\cos[n(\phi-\Psi_{n})].
\end{equation}
\end{linenomath}
\noindent Here, the Fourier coefficients, $v_{n}$, characterize the strength of
the anisotropic flow, while the azimuthal flow orientation is represented
by the corresponding ``event plane'' angle, $\Psi_n$, the direction of maximum final-state particle density.
The event plane angles
are related to the event-by-event spacial distribution of the participating
nucleons in the initial overlap region. The most widely studied and typically also strongest form of
anisotropic flow is the second Fourier component, $v_2$, called ``elliptic
flow''. The elliptic flow event plane, $\Psi_2$, is correlated with the ``participant plane''
given by the beam direction and the shorter axis of the approximately elliptical nucleon overlap region.
Because of event-by-event fluctuations, higher-order deformations or eccentricities
of the initial geometry can also be induced, which lead to higher-order Fourier
harmonics ($v_n$, $n \ge 3$) in the final state with respect to their corresponding event plane angles,
$\Psi_n$~\cite{Alver:2010gr}.
Studies of azimuthal anisotropy harmonics provide important information on
the fundamental transport properties of the medium, e.g., the ratio of shear
viscosity to entropy density, $\eta/s$ ~\cite{Alver:2010dn,Schenke:2010rr,Qiu:2011hf}.

A commonly used experimental method to determine the single-particle
azimuthal
anisotropy harmonics, $v_n$, is the measurement of two-particle
azimuthal correlations~\cite{Danielewicz:1985hn,Ollitrault:1993ba,Voloshin:1994mz,Poskanzer:1998yz}.
The azimuthal distribution of particle pairs as a function of
their relative azimuthal angle $\dphi$ can also be characterized by its Fourier components,
\begin{linenomath}
\begin{equation}
\label{eq:fourier_two}
\frac{\rd{}N^\text{pair}}{\rd\dphi} \propto 1+2\sum_{n} V_{n\Delta}\cos(n\dphi).
\end{equation}
\end{linenomath}
If the dominant source of final-state particle correlations is
collective flow,
the two-particle Fourier coefficients, $V_{n\Delta}$, are commonly expected to follow
the factorization relation:
\begin{equation}
\label{eq:fact}
V_{n\Delta}=v^{a}_{n}\; v^{b}_{n},
\end{equation}
where $v^{a}_{n}$ and $v^{b}_{n}$ represent the single-particle anisotropy
harmonics for a pair of particles ($a$ and $b$) in the event.
The particle pairs can be selected from the same or
different transverse momentum (\pt) and pseudorapidity ($\eta$) ranges.
Here, a key assumption is that the event plane angle $\Psi_n$ in Eq.~(\ref{eq:fourier_single})
is a global phase angle for all particles of the entire event, which is canceled
when taking the azimuthal angle difference between two particles. As a result,
the flow-driven $\dphi$ distribution in Eq.~(\ref{eq:fourier_two}) has no
dependence on $\Psi_n$.
The most common approach to obtain the
single-particle $v_n$ in the two-particle method is to fix one particle in a wide \pt ($\eta$) region and measure $V_{n\Delta}$ by only varying \pt ($\eta$) of the
other particle to determine $v_n$ as a function of \pt ($\eta$).

However, a significant breakdown of the factorization assumption, up to about 20\%,
was recently observed for pairs of particles, separated by more than 2
units in $\eta$, from different \pt ranges in ultra-central (0--0.2\% centrality) \PbPb collisions~\cite{CMS:2013bza}. The centrality in heavy ion collisions is defined as a fraction
of the total inelastic \PbPb cross section, with 0\% denoting the most central collisions.
While nonflow correlations (such as back-to-back jets)
have been speculated to possibly account for this effect,
contributions of those short-range correlations to the collective anisotropy are
less dominant in high-multiplicity events as the total number of particles
increases~\cite{Chatrchyan:2013nka}.
It was then realized that in hydrodynamic models the assumption of factorization
does not hold in general because of fluctuations in the initial overlap region of two nuclei~\cite{Gardim:2012im,Heinz:2013bua}.
In each event, due to local perturbations in the energy density distribution generating a pressure
gradient that drives particles in random directions with differing boosts, the resulting event plane angles found with
final-state particles from different \pt ranges may fluctuate with respect to each other
(although still correlated with the initial participant plane). This effect of
initial-state fluctuations thus breaks the factorization relation of
Eq.~(\ref{eq:fact}), which assumes a unique event plane angle for all particles in an event.
As a result, the precise meaning of previous single-particle $v_n$ results
should be reinterpreted as being
with respect to the event plane determined with particles over a specific, usually wide, \pt range.
Quantitative studies of the factorization breakdown effect as a function of \pt could place stringent constraints
on the spatial scale (or granularity) of the fluctuations in the initial
state of heavy ion collisions, especially along the radial direction~\cite{Kozlov:2014fqa,Floerchinger:2013rya,ColemanSmith:2012ka}.

The recent observation of long-range near-side ($\dphi \sim 0$)
two-particle correlations
in \pp \cite{Khachatryan:2010gv} and \pPb \cite{CMS:2012qk,alice:2012qe,Aad:2012gla}
collisions raised the question of whether hydrodynamic flow is developed
also in these small collision systems. The extracted
$v_n$ harmonics in \pPb collisions have been studied in detail
as a function of \pt and event multiplicity~\cite{Chatrchyan:2013nka,Aad:2014lta}.
The initial-state geometry of a \pPb collision is expected to be entirely
driven by fluctuations. If the observed long-range correlations in such collisions
indeed originate from hydrodynamic flow, the effect of factorization
breakdown should also be observed in the data and described by
hydrodynamic models. Since the initial-state geometries of both high-multiplicity \pPb and ultra-central \PbPb collisions
are dominated by fluctuations, it is of great interest to investigate
whether the magnitude of factorization breakdown is similar in these two systems.

Furthermore, the factorization breakdown in $\eta$ is sensitive to event plane fluctuations at different
$\eta$~\cite{Gardim:2012im}. This phenomenon has been investigated in hydrodynamic and
parton transport models~\cite{Bozek:2010vz,Pang:2014pxa,Xiao:2012uw,Jia:2014vja}.
The observation and study of this effect will provide new insights into the dynamics
of longitudinal expansion of the hot quark and gluon medium, and serves as an ideal test ground of three-dimensional hydrodynamic models.

This paper presents a comprehensive investigation of the factorization breakdown effect in
two-particle azimuthal Fourier harmonics in \PbPb (\pPb) collisions
at $\rootsNN = 2.76\;(5.02)$\TeV, to search for evidence of \pt- and
$\eta$-dependent event plane fluctuations.
The Fourier harmonics of two-particle azimuthal correlations are extracted for pairs
with $\abs{\deta}>2$ as a function of \pt and $\eta$ of both particles in a pair.
The results are presented over a wide range of centrality or
event multiplicity classes, and are compared with hydrodynamic models
in \PbPb and \pPb collisions. As the \pt- and $\eta$-dependent aspects
of factorization breakdown probe system dynamics in the transverse and longitudinal directions, respectively,
an assumption is made that the dependence on each variable can be studied independently
by averaging over the other, and two different analysis techniques are applied.
These two aspects of the analysis are described in Sections~\ref{sec:pt}
and \ref{sec:eta} separately, including the analysis procedures and results.

\section{Experimental setup and data sample}
\label{sec:exp}

A comprehensive description of the Compact Muon Solenoid (CMS) detector at 
the CERN LHC, together
with a definition of the coordinate system used and the relevant kinematic
variables, can be found in Ref.~\cite{JINST}. The main detector sub-system
used in this paper is the tracker, located in a superconducting solenoid
of 6\unit{m} internal diameter, providing a magnetic field of 3.8\unit{T}.
The tracker consists of 1440 silicon pixel and 15\,148 silicon strip detector
modules, covering the pseudorapidity range $\abs{\eta}<2.5$. For hadrons with $\pt
\approx 1\GeVc$ and $\abs{\eta} \approx 0$, the impact parameter resolution is
approximately 100\mum and the \pt resolution is 0.8\%.

The electromagnetic calorimeter (ECAL) and the hadron calorimeter (HCAL) are
also located inside the solenoid. The ECAL consists of 75\,848
lead tungstate crystals, arranged in a quasi-projective geometry and distributed in a
barrel region ($\abs{\eta} < 1.48$) and two endcaps that extend to $\abs{\eta} = 3.0$.
The HCAL barrel and endcaps are sampling calorimeters composed of brass and
scintillator plates, covering $\abs{\eta} < 3.0$. In addition, CMS has an
extensive forward calorimetry, in
particular two steel/quartz-fiber Cherenkov hadronic forward (HF)
calorimeters, which cover the pseudorapidity range $2.9 < \abs{\eta} < 5.2$.
The HF calorimeters are segmented into towers, each of which is a two-dimensional
cell with a granularity of 0.5 in $\eta$ and 0.349 radians in $\phi$.
A set of scintillator tiles, the beam scintillator counters (BSC), are mounted
on the inner side of the HF calorimeters and are used for triggering and beam-halo rejection.
The BSCs cover the range $3.23 < \abs{\eta} < 4.65$.
The detailed Monte Carlo (MC) simulation of the CMS detector response is
based on \GEANTfour \cite{GEANT4}.

The data sample used in this analysis was collected with the CMS detector during
the LHC \PbPb run in 2011 and \pPb run in 2013. The total integrated luminosity of
the data sets is about 159\mubinv\ for \PbPb, and 35\nbinv
for \pPb.
During the \pPb run, the beam energies were 4\TeV
for protons and 1.58\TeV per nucleon for lead nuclei, resulting in a center-of-mass
energy per nucleon pair of 5.02\TeV. As a result of the energy
difference between the colliding beams, the nucleon-nucleon center-of-mass in the \pPb collisions is not at rest in the laboratory frame. Massless particles
emitted at $\eta_\mathrm{cm} = 0$ in the nucleon-nucleon center-of-mass frame
will be detected at $\eta = -0.465$ or 0.465 (clockwise or counterclockwise
proton beam) in the laboratory frame.

\section{Selection of events and tracks}

Online triggers, offline event selections, track reconstruction and
selections are identical to those used in previous analyses of \PbPb and
\pPb\ data~\cite{Chatrchyan:2013nka,CMS:2013bza} and are briefly outlined in the following sections.

\subsection{\PbPb\ data}
\label{subsec:PbPbdata}
Minimum bias \PbPb events were selected using coincident trigger signals from both
ends of the detector in either BSCs or the HF calorimeters. Events due to detector noise,
cosmic rays, out-of-time triggers, and beam backgrounds were suppressed
by requiring a coincidence of the minimum bias trigger with bunches colliding
in the interaction region. The trigger has an efficiency of $(97 \pm 3)$\%
for hadronic inelastic \PbPb collisions. Because of hardware limits on the data
acquisition rate, only a small fraction (2\%) of all minimum bias
events were recorded (i.e., the trigger is ``prescaled'').
To enhance the event sample for very central \PbPb collisions, a dedicated online
trigger was implemented by simultaneously requiring the HF
transverse energy (\ET) sum to be greater than 3260\GeV and the pixel cluster
multiplicity to be greater than 51400 (which approximately corresponds to 9500 charged particles
over 5 units of pseudorapidity). The selected events correspond
to the 0.2\% most central \PbPb collisions.
Other standard \PbPb centrality classes presented in this paper are determined based
on the total energy deposited in the HF calorimeters~\cite{Chatrchyan:2012ta}.
The inefficiencies of the minimum bias trigger and event
selection for very peripheral events are properly taken into account.

To further reduce the background from single-beam interactions (e.g., beam-gas and beam-halo),
cosmic muons, and ultra peripheral collisions that lead to the electromagnetic breakup of one or
both \Pb nuclei~\cite{Djuvsland:2010qs}, offline \PbPb event selection criteria~\cite{Chatrchyan:2012ta} are applied
by requiring energy deposits in at least three towers in each of the HF calorimeters, with
at least 3\GeV of energy in each tower, and the presence of a reconstructed
primary vertex containing at least two tracks. The reconstructed
primary vertex is required to be located within $\pm$15\unit{cm} of the average
interaction region along the beam axis and within a radius of 0.02\unit{cm}
in the transverse plane. Following the procedure developed in Ref.~\cite{CMS:2013bza},
events with large signals in both Zero Degree Calorimeter (ZDC) and HF
are identified as having at least one additional interaction, or pileup events, and thus rejected (about 0.1\% of all events).

The reconstruction of the primary event vertex and of the trajectories of
charged particles in \PbPb collisions are based on signals in the silicon
pixel and strip detectors and described in detail in
Ref.~\cite{Chatrchyan:2012ta}. From studies based on \PbPb events
simulated using \textsc{hydjet} v1.8~\cite{Lokhtin:2005px}, the
combined geometrical acceptance and reconstruction efficiency of the
primary tracks is about 70\% at $\pt\sim1\GeVc$ and $\abs{\eta} < 1.0$ for
the most central 0--5\% \PbPb events, but drops to about 50\% for $\pt \sim 0.3$\GeVc.
The fraction of misidentified tracks is kept at the level of $<$5\% over
most of the \pt ($>$0.5\GeVc) and $\abs{\eta}$ ($<$1.6) ranges.
It increases to about 20\% for very low \pt ($<$0.5\GeVc) particles in
the forward ($\abs{\eta} \ge 2)$ region.

\subsection{\pPb\ data}

Minimum bias \pPb events were selected by requiring that at least
one track with $\pt > 0.4$\GeVc is found in the pixel tracker
in coincidence with a \pPb bunch crossing. About 0.1\% of all minimum bias \pPb events were recorded.
In order to select high-multiplicity \pPb collisions,
a dedicated high-multiplicity trigger was implemented using the CMS level-1
(L1) and high-level trigger (HLT) systems. At L1, the total transverse energy
measured using both ECAL and HCAL is required to be greater than a given threshold (20 or 40\GeV).
Online track reconstruction for the HLT was based on
the three layers of pixel detectors, and required a track origin within a cylindrical region,
centered at the average interaction point of two beams, of
length 30\unit{cm} along the beam and radius 0.2\unit{cm} perpendicular to the beam.
For each event, the vertex reconstructed with the highest number of pixel tracks was selected.
The number of pixel tracks (${N}_\mathrm{trk}^\mathrm{online}$)
with $\abs{\eta}<2.4$, $\pt > 0.4\GeVc$, and a distance of closest approach of 0.4\unit{cm} or
less to this vertex, was determined for each event.

Offline selections similar to those used for the \PbPb data sample are applied
to reject non-hadronic \pPb interactions.
A coincidence of at least one HF calorimeter tower with more than 3\GeV of
total energy in each of the HF detectors is required.
Events are also required to contain at least one reconstructed primary
vertex within 15\unit{cm} of the nominal interaction point along the beam axis
and within 0.15\unit{cm} transverse to the beam trajectory.
At least two reconstructed tracks are required to be associated
with the primary vertex. Beam-related background is suppressed
by rejecting events for which less than 25\% of all reconstructed tracks are
of  sufficiently good quality to be tracks selected for physics analysis,
as will be discussed later in this section.
Among those \pPb interactions simulated with the \EPOS~\cite{Porteboeuf:2010um}
and \HIJING~\cite{Gyulassy:1994ew} event generators that have at least one primary
particle with total energy $E>3$\GeV in both $\eta$ ranges of $-5<\eta<-3$ and
$3<\eta<5$, the above criteria are found to select 97--98\% of the events.
Pileup events are removed based on the number of tracks
associated with each vertex in a bunch crossing and the distance between
different vertices~\cite{Chatrchyan:2013nka}.
A purity of 99.8\% for single \pPb collision events is achieved for the highest
multiplicity \pPb interactions studied in this paper.

For the \pPb analysis, the standard track reconstruction as in \pp\
collisions is applied.
The CMS high-purity tracks (as defined in Ref.~\cite{Chatrchyan:2014fea}) are used.
Additionally, a reconstructed track is only considered as a
primary-track candidate if the significance of the separation along the beam axis ($z$)
between the track and primary vertex, $d_z/\sigma(d_z)$, and the significance
of the impact parameter relative to the primary vertex transverse to the beam,
$d_{\rm T}/\sigma(d_{\rm T})$, are each less than 3. The relative uncertainty
in the transverse momentum measurement, $\sigma(\pt)/\pt$, is required to
be less than 10\%.
To ensure high tracking efficiency and to reduce the rate of misidentified tracks,
only tracks within $\abs{\eta}<2.4$ and with $\pt >$ 0.3\GeVc are used in the analysis.

The entire \pPb data set is divided into classes of reconstructed track multiplicity, \noff,
where primary tracks with $\abs{\eta}<2.4$ and $\pt >0.4$\GeVc are counted.
The multiplicity classification in this analysis is identical to that used in Ref.~\cite{Chatrchyan:2013nka},
where more details are provided. The more central (0--50\%) \PbPb data, including
ultra-central triggered events, are analyzed with a standard reconstruction algorithm
used in heavy ion collisions, as described in Section \ref{subsec:PbPbdata}.
In order
to compare the \pPb and \PbPb systems at the same collision multiplicity,
peripheral \PbPb events for 50--100\% centrality are reprocessed using the same event
selections and track reconstruction as for the \pPb analysis.

\section{Transverse momentum dependence of factorization breakdown}
\label{sec:pt}

\subsection{Analysis technique}
\label{subsec:ana_pt}

The \pt-dependent factorization breaking effect is investigated using the same
analysis technique of two-particle azimuthal correlations
as that applied in Ref.~\cite{CMS:2013bza}. For simplicity, a pair
of two charged tracks are labeled as particle $a$ and $b$ (equivalent to
the trigger and associated
particles used in previous publications). They are selected
from the same or different \pttrg and \ptass ranges within $\abs{\eta^{a,b}}<2.4$.
The two-particle Fourier coefficients, $V_{n\Delta}$, are calculated as the average value
of $\cos(n\Delta\phi)$ over all particle pairs, which fulfill the requirement of $ \abs{\deta} > 2$ (to avoid the
short-range correlations from jets and resonance decays):

\begin{equation}
\label{average}
V_{n\Delta} \equiv \avg{\cos(n\dphi)}_{S}-\avg{\cos(n\dphi)}_{B},
\end{equation}

\noindent in given ranges of \pttrg and \ptass. Here, $\avg{\ }$ denotes
averaging over all particle pairs in each event and over all the events. The subscript
$S$ corresponds to the average over pairs taken from the same event, while $B$
represents the mixing of particles from two randomly-selected events in the same 2\unit{cm} wide range
of the primary vertex position in the $z$ direction and from the same centrality (track multiplicity) class.
The $\avg{\cos(n\Delta\phi)}_{B}$ term, which is typically two orders of
magnitude smaller than the corresponding $S$ term, is
subtracted to account for the effects of detector non-uniformity. This analysis
is equivalent to those in Refs.~\cite{Chatrchyan:2011eka,Chatrchyan:2012wg,Chatrchyan:2013nka,Khachatryan:2014jra},
where the two-particle azimuthal correlation function is first constructed
and then fit with
 a Fourier series. The advantage of the present approach is that the extracted Fourier
harmonics will not be affected by the finite bin widths of the histogram in $\Delta\eta$
and $\Delta\phi$ of the two-particle correlation function, which is relevant for higher-order
Fourier harmonics.

With the $V_{n\Delta}(\pttrg,\ptass)$ values as a function of \pttrg and \ptass,
the factorization ratio,
\begin{linenomath}
\begin{equation}
    \label{r_n_def}
    r_{n}(\pttrg,\ptass) \equiv \frac{V_{n\Delta}(\pttrg,\ptass)}{\sqrt{V_{n\Delta}(\pttrg,\pttrg)V_{n\Delta}(\ptass,\ptass)}},
\end{equation}
\end{linenomath}
has been proposed as a direct measurement of the factorization
breakdown effect and
to explore the \pt-dependent event plane angle fluctuations in the context of hydrodynamics~\cite{Gardim:2012im}.
Here, the $V_{n\Delta}$ coefficients are calculated by pairing particles
within the same \pt interval (denominator) or from different \pt intervals (numerator). If the factorization relation of
Eq.~(\ref{eq:fact}) holds, this ratio is expected to be unity. However,
with the presence of a \pt-dependent event plane angle,
it can be shown that the factorization ratio, $r_n$, is equivalent to
\begin{linenomath}
\begin{equation}
    \label{r_n}
    r_{n}(\pttrg,\ptass) =\frac{\langle 
v_{n}(\pttrg)v_{n}(\ptass)\cos\{n\big[\Psi_{n}(\pttrg)-\Psi_{n}(\ptass)\big]\} 
\rangle}{\sqrt{\langle v_{n}^{2}(\pttrg) \rangle\langle v_{n}^{2}(\ptass) \rangle}},
\end{equation}
\end{linenomath}
where $\Psi_{n}(\pttrg)$ and $\Psi_{n}(\ptass)$ represent
the event plane angles determined using particles
from \pttrg and \ptass\ intervals, respectively~\cite{Gardim:2012im,Heinz:2013bua},
and $\langle \rangle$ denotes averaging over all the events.
As one can see from Eq.~(\ref{r_n}), $r_n$ is in general
less than unity in the presence of the \pt-dependent event plane angle $\Psi_{n}$ fluctuations.

\subsection{Results for \PbPb data}

\begin{figure*}[htb]
\centering
    \includegraphics[width=\linewidth]{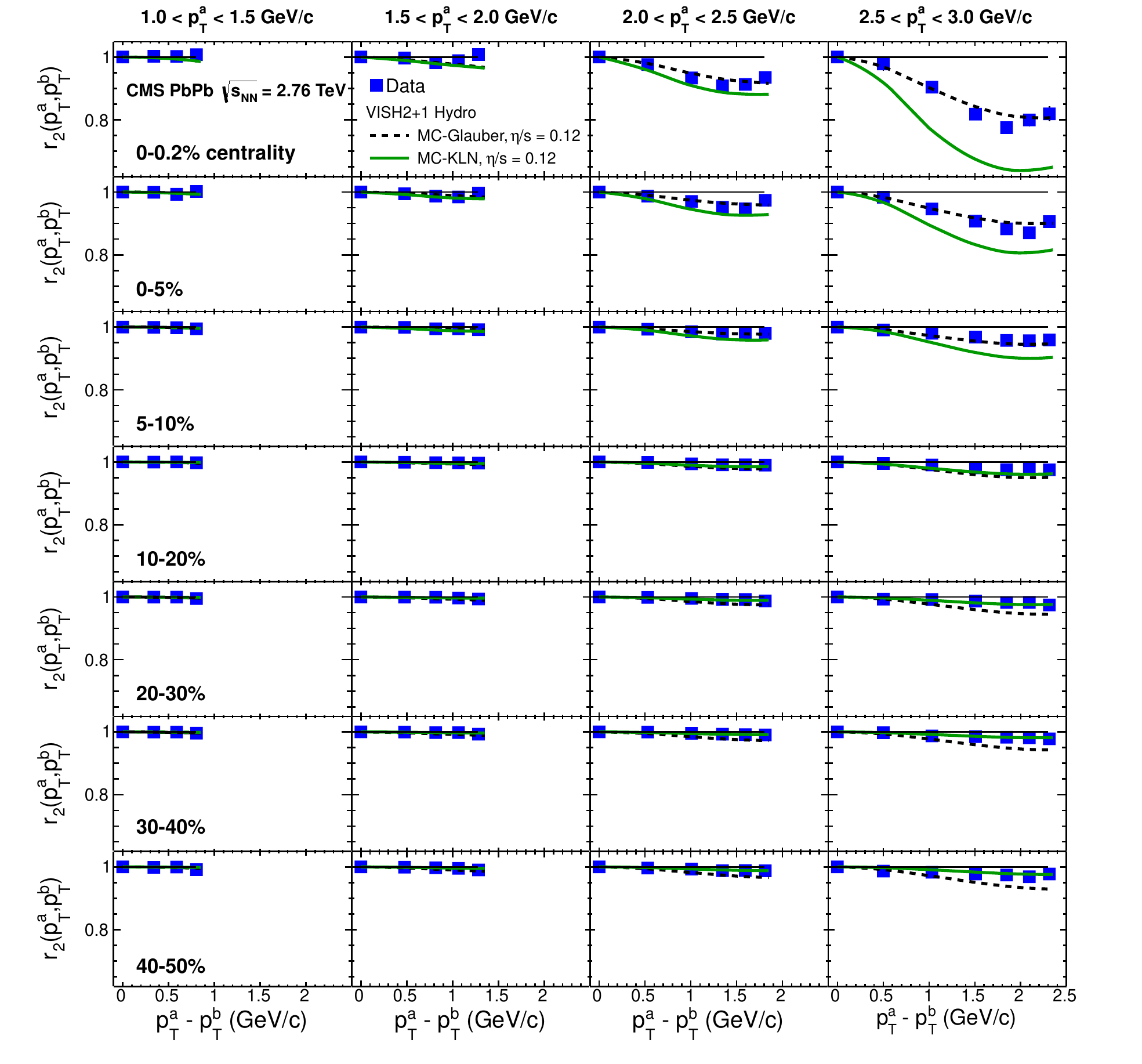}
    \caption{(Color online) The \pt-dependent factorization ratio, $r_2$,
as a function of $\pttrg-\ptass$ in bins
    of \pttrg for different centrality ranges of \PbPb collisions at $\rootsNN = 2.76$\TeV.
    The curves show the calculations from a viscous hydrodynamic model~\cite{Heinz:2013bua}
    using MC-Glauber and MC-KLN initial condition models, and an $\eta/s$ value of 0.12.
    Each row represents a different centrality range, while each column corresponds to a
    different \pttrg range. The horizontal solid lines denote the $r_{2}$ value of unity.
    The error bars correspond to statistical uncertainties, while systematic
    uncertainties are negligible for the $r_n$ results, and thus are not shown.
    }
    \label{fig:r2_PbPb}
\end{figure*}

\begin{figure*}[htb]
\centering
    \includegraphics[width=\linewidth]{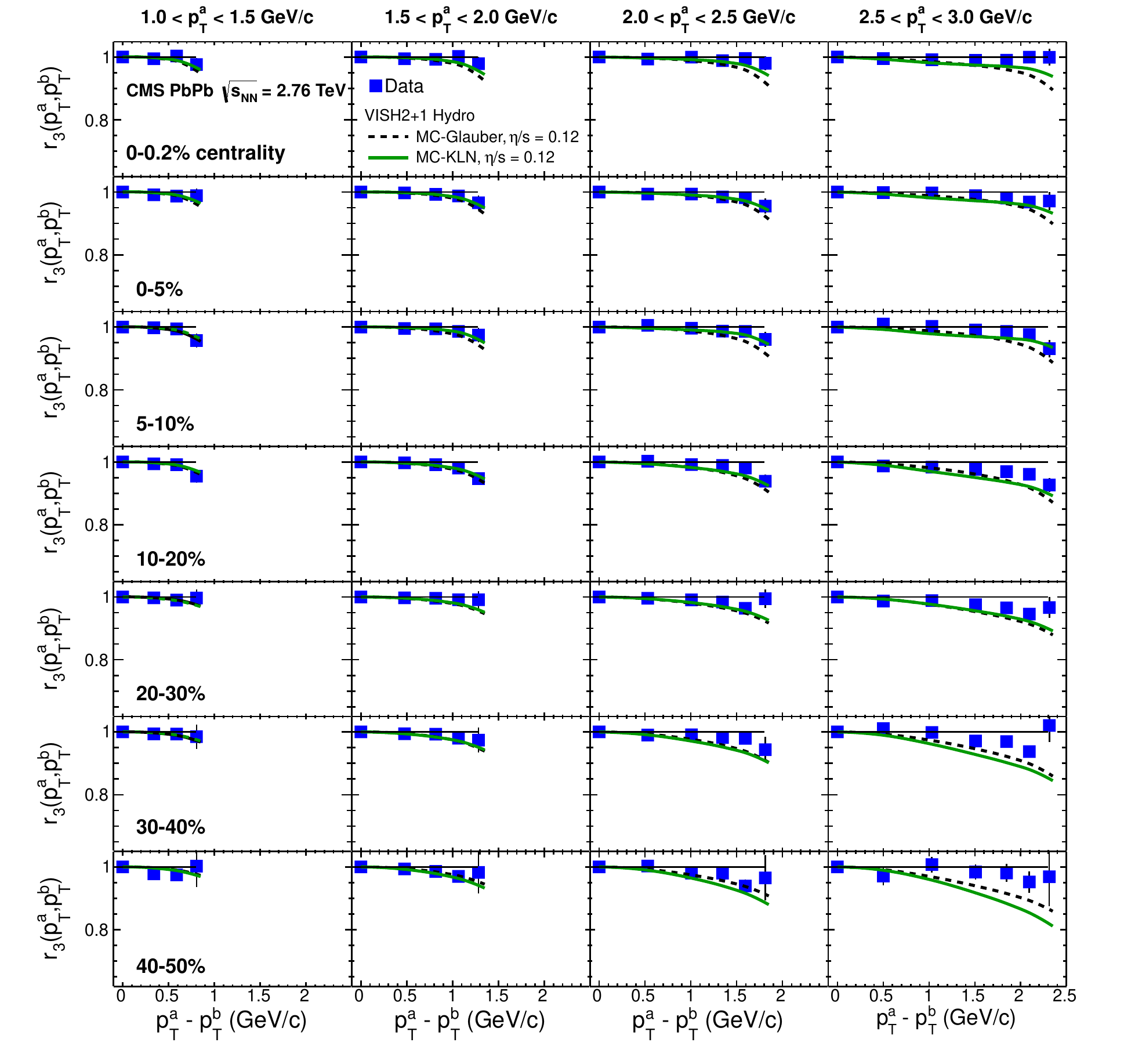}
    \caption{(Color online) Similar distributions as shown in
Fig.~\ref{fig:r2_PbPb},
but for the factorization ratio $r_3$.
    }
    \label{fig:r3_PbPb}
\end{figure*}

\begin{figure*}[h!tb]
\centering
    \includegraphics[width=\linewidth]{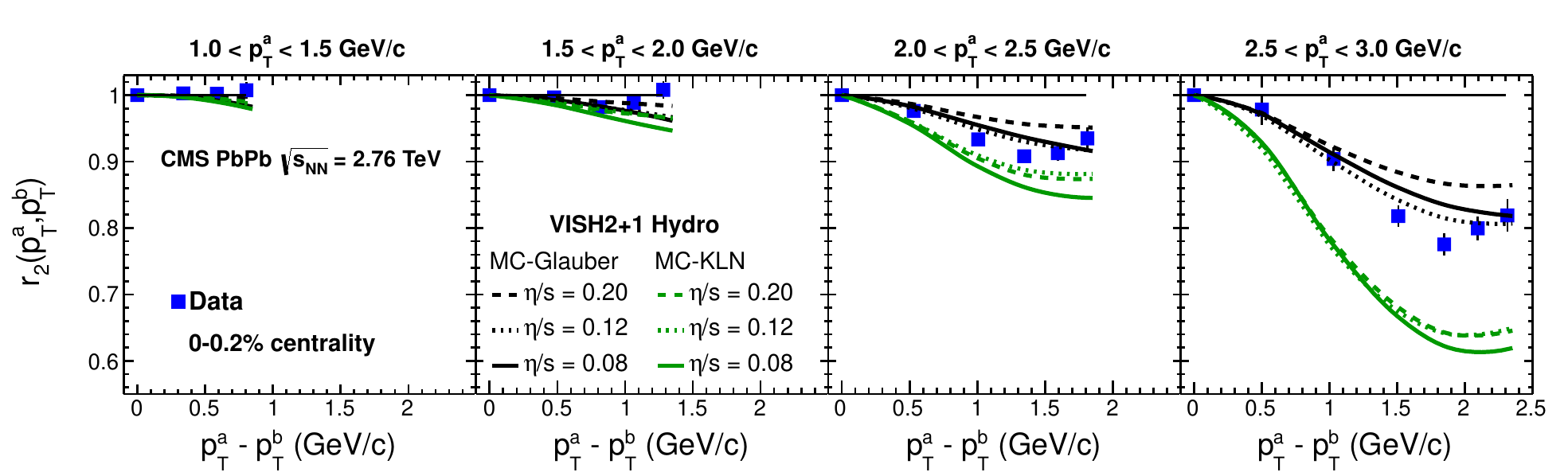}
    \caption{(Color online) Factorization ratio, $r_2$, as a function of
$\pttrg-\ptass$ in bins
    of \pttrg for 0--0.2\% centrality \PbPb data at $\rootsNN = 2.76$\TeV
    compared to viscous hydrodynamic calculations~\cite{Heinz:2013bua} using MC-Glauber
    and MC-KLN initial condition models, and three different values of $\eta/s$.
    The horizontal solid lines denote the $r_{2}$ value of unity. The error
bars correspond to statistical uncertainties, while systematic
    uncertainties are negligible for the $r_n$ results and thus are not shown.
    }
    \label{fig:rn_PbPbUCC}
\end{figure*}

The first measurement of \pt-dependent factorization breakdown in \PbPb collisions
was presented in Ref.~\cite{CMS:2013bza}. Our analysis is expanded to cover
a much wider centrality range from 0\% to 50\%, and also includes a systematic comparison to hydrodynamic models.
The values of $r_{2}(\pttrg,\ptass)$ and $r_{3}(\pttrg,\ptass)$ in \PbPb collisions at $\rootsNN = 2.76$\TeV are presented as a function of $\pttrg-\ptass$
in Figs.~\ref{fig:r2_PbPb} and \ref{fig:r3_PbPb}, for several \pttrg ranges
in seven different centrality classes from 0--0.2\% to 40--50\%.
The average \pt values within each \pttrg and \ptass range are used in order
to calculate the difference between \pttrg and \ptass.
By construction, the $r_{n}$ value for the highest analyzed \ptass range,
where both particles are selected from the same \pt
interval, is equal to one. Only results for $\pttrg \geq \ptass$ are presented, with a maximal \pttrg value of 3\GeVc,
a kinematic regime where the hydrodynamic flow effect is believed to be dominant.
The error bars correspond to statistical uncertainties, while systematic
uncertainties are found to be negligible for the $r_n$ results (mainly because
systematic uncertainties of $V_{n\Delta}$ are typically on the order of a few percent, and ratios
of $V_{n\Delta}$ are taken to form $r_n$ in this paper, where systematic uncertainties mostly cancel) and thus are not shown
in any of the figures.

A clear deviation from unity of the $r_2$ value (Fig.~\ref{fig:r2_PbPb}) is
observed for the highest \pt ranges in very central PbPb collisions. For
each centrality class, the effect
becomes more pronounced with an increase of \pttrg and also
the difference between \pttrg and \ptass values.
This trend is expected as event-by-event initial-state geometry fluctuations
play a more dominant role as the collisions become more central.
The factorization breakdown effect reaches 20\%
in the ultra-central 0--0.2\% events for the greatest difference between
\pttrg and \ptass. For more peripheral centrality
classes, the maximum effect is a few percent.
Calculations using viscous hydrodynamics~\cite{Heinz:2013bua}
are performed in all centrality classes, and shown as the curves in Fig.~\ref{fig:r2_PbPb}.
To focus on the effect of initial-state fluctuations, the $\eta/s$ value is fixed at 0.12.
Two different models of initial conditions, MC-Glauber~\cite{glauber,Alver:Glauber}
and MC-Kharzeev–Levin–Nardi (MC-KLN; motivated by the concept of gluon 
saturation)~\cite{Drescher:2006pi}, are compared to data. The qualitative
trend of
the data is consistent with hydrodynamic calculations. However, quantitatively,
neither of the two models can describe all the data. The MC-Glauber model matches better
the data for central collisions, while MC-KLN model appears to describe the data
in the peripheral centrality range.

For the third-order harmonics ($n=3$), the effect of factorization breakdown is
significantly smaller than for the second-order harmonics.
Only a weak centrality dependence of $r_3$ is seen in Fig.~\ref{fig:r3_PbPb}.
The biggest deviation of $r_3$ from unity is about 5\% at large values of $\pttrg-\ptass$
(i.e., $> 1$\GeVc). Again, the qualitative features of the data are described by
the hydrodynamic model, although the effects are over-estimated for peripheral collisions by the model.
Calculations of $r_3$ using two different initial-state models yield similar
results, with MC-KLN model showing a slightly stronger centrality dependence.

To understand better how the effects of factorization breakdown and
\pt-dependent event plane fluctuations are influenced by
the initial-state conditions and the value of $\eta/s$ in hydrodynamic
models, a detailed comparison of measured $r_2$ values
in 0--0.2\% centrality PbPb collisions (where the effect is most evident) to hydrodynamic calculations is shown in
Fig.~\ref{fig:rn_PbPbUCC}.
For this comparison, calculations with MC-Glauber and MC-KLN initial conditions
are each performed for three different $\eta/s$ values and compared to data.
For each initial-state model, the $r_2$ values
are found to be largely insensitive to different values of $\eta/s$. This is because,
in defining $r_{n}(\pttrg,\ptass)$, the magnitudes of anisotropy harmonics, which have
a much greater sensitivity to $\eta/s$, are mostly canceled. Fluctuations of the event plane angle
in \pt are mainly driven by the non-smooth local fluctuations in the initial energy density distribution.
This comparison shows that the use of $r_n$ data can provide new constraints on the detailed modeling
of the initial-state condition and the fluctuations of the medium created in
heavy ion collisions, which is independent of the $\eta/s$ value.
The better constraints on the initial-state conditions found using the
$r_n$ data will, in turn,
improve the uncertainties of determining the medium's transport properties (e.g., $\eta/s$)
using other experimental observables (e.g., the $v_n$ magnitude, which is sensitive to both the initial state and $\eta/s$).

\subsection{Results for \pPb data}

\begin{figure*}[t!h]
\centering
    \includegraphics[width=\linewidth]{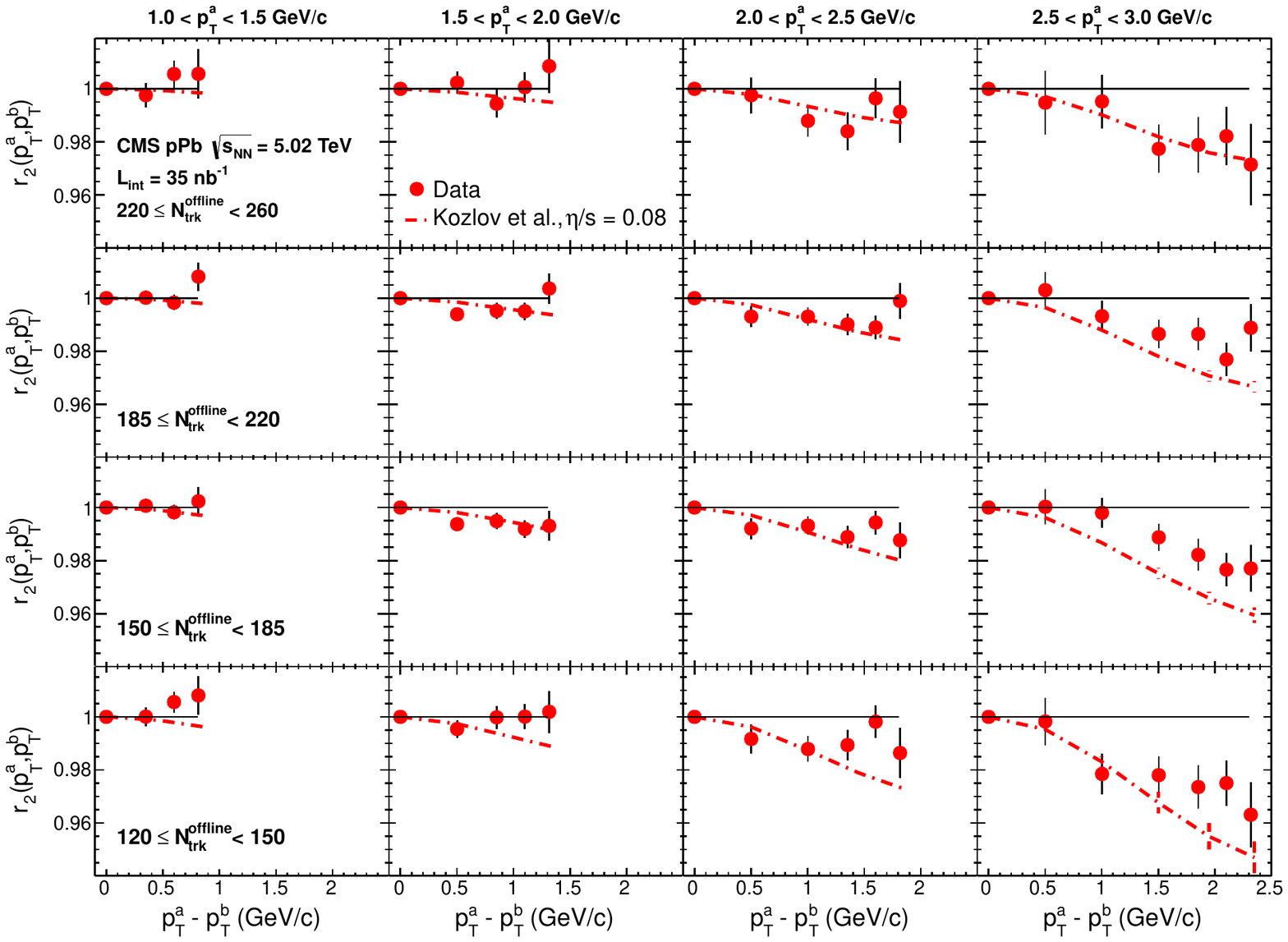}
    \caption{(Color online) The \pt-dependent factorization ratio, $r_2$,
as a function of
    $\pttrg -\ptass$ in bins of \pttrg for four \noff\ ranges in 5.02\TeV \pPb collisions.
    The curves show the predictions from hydrodynamic calculations for
\pPb\ collisions of Ref.~\cite{Kozlov:2014fqa}. The horizontal solid lines denote the $r_{2}$ value of unity.
    The error bars correspond to statistical uncertainties, while systematic
    uncertainties are negligible for the $r_n$ results and thus are not shown.
    }
    \label{fig:r2_pPb}
\end{figure*}

\begin{figure*}[t!h]
\centering
    \includegraphics[width=\linewidth]{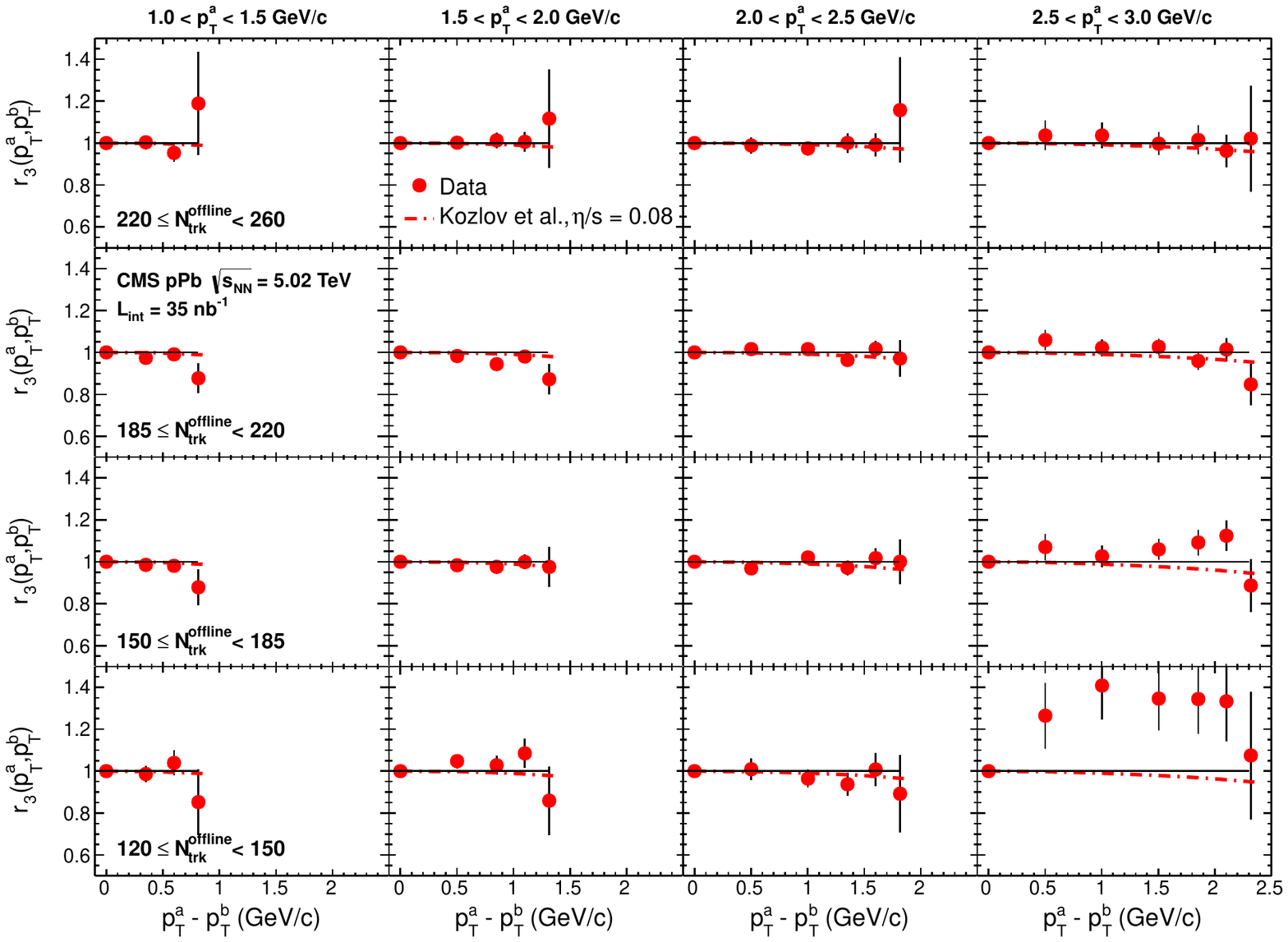}
    \caption{(Color online) Similar distributions as shown in
Fig.~\ref{fig:r2_pPb}, but
for the factorization ratio $r_3$.
    }
    \label{fig:r3_pPb}
\end{figure*}

To gain insights into the origin of long-range correlations observed in high-multiplicity \pPb collisions,
the measurement of $r_2$ and $r_3$ is also performed for \pPb data at $\rootsNN = 5.02\TeV$
for four different high-multiplicity ranges. The results are shown in Figs.~\ref{fig:r2_pPb}
and \ref{fig:r3_pPb}, in the same format as those for \PbPb collisions,
for four \pttrg ranges (of increasing \pt from left to right panels) as a function of  $\pttrg - \ptass$.

Breakdown of factorization is observed in the $r_2$ results of \pPb collisions
for all multiplicity ranges investigated in this paper. Similar to \PbPb collisions, for any multiplicity range, the
effect gets larger with an increase in the difference between \pttrg and \ptass values.
However, the observed factorization breakdown reaches only up to 2--3\% for the largest value of $\pttrg-\ptass$ at
$2.5<\pttrg<3.0$\GeVc. This is significantly smaller than that seen in central \PbPb collisions. Little multiplicity dependence of $r_2$ is observed in \pPb collisions.
Comparison of the CMS data to hydrodynamic predictions
for \pPb collisions in Ref.~\cite{Kozlov:2014fqa} is also shown. In this hydrodynamic calculation, a modified MC-Glauber initial-state
model is employed for pPb collisions where the contributing entropy density of each participating
nucleon in the transverse plane is distributed according to a 2D Gaussian distribution.
The width of the transverse Gaussian function can be chosen to vary the
transverse granularity
of fluctuations, to which the $r_n$ values are found to be most sensitive.
The $r_2$ data are better described by calculations with a width parameter of 0.4\unit{fm} (curves in Fig.~\ref{fig:r2_pPb}),
while a width of 0.8~fm gives an $r_n$ value of nearly unity (not shown) and thus underestimates
the effect observed in the data. For both cases, the calculations are found
to be insensitive to different $\eta/s$ values, consistent with the hydrodynamic calculations
used for more central \PbPb collisions presented earlier.

Results of $r_3$ are shown in Fig.~\ref{fig:r3_pPb},
presented in the same format as for $r_2$. Within current statistical precision,
no evident breakdown of factorization is found in very-high-multiplicity \pPb events
($185<\noff<260$), while the $r_3$ value exceeds unity for much lower-multiplicity
\pPb\ events at high \pt, particularly for $120<\noff<150$. This is a clear
indication of significant nonflow effects as the event multiplicity decreases,
because the $r_n$ values predicted by hydrodynamic models with \pt-dependent
event plane angle fluctuations would always be equal to or less than one,
according to Eq.~(\ref{r_n}). One obvious possibility is back-to-back jet correlations,
which would give a large negative contribution to $V_{3\Delta}$ at high
\pttrg and \ptass values in low multiplicity events~\cite{Chatrchyan:2012wg}.
This would lead to a significant reduction of
the denominator of Eq.~(\ref{r_n}) and drives the $r_3$ value up above unity.
Very little effect of factorization breakdown for $n=3$ is predicted in
Ref.~\cite{Kozlov:2014fqa}, which is consistent with the data except
for the low-multiplicity ranges.

\subsection{Comparison of \pPb and \PbPb data}

\begin{figure*}[ht]
\centering
    \includegraphics[width=\linewidth]{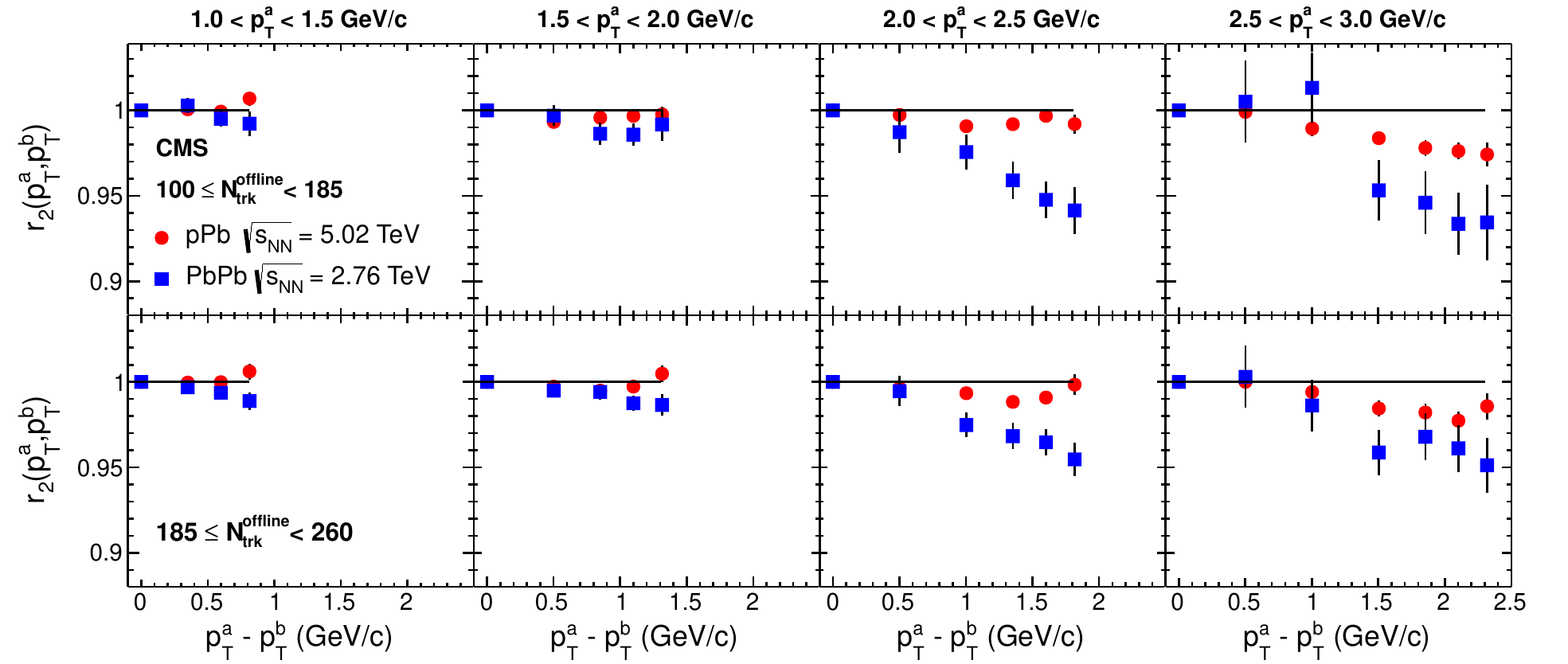}
    \caption{(Color online) The \pt-dependent factorization ratio, $r_2$,
as a function of $\pttrg - \ptass$ in bins
    of \pttrg for two \noff\ ranges of 5.02\TeV \pPb and 2.76\TeV \PbPb collisions.
    The horizontal solid lines denote the $r_{2}$ value of unity. The error
bars correspond to statistical uncertainties, while systematic
    uncertainties are negligible for the $r_n$ results and thus are not shown.
    }
    \label{fig:rn_pPb}
\end{figure*}

\begin{figure*}[th]
\centering
    \includegraphics[width=\cmsFigWidth]{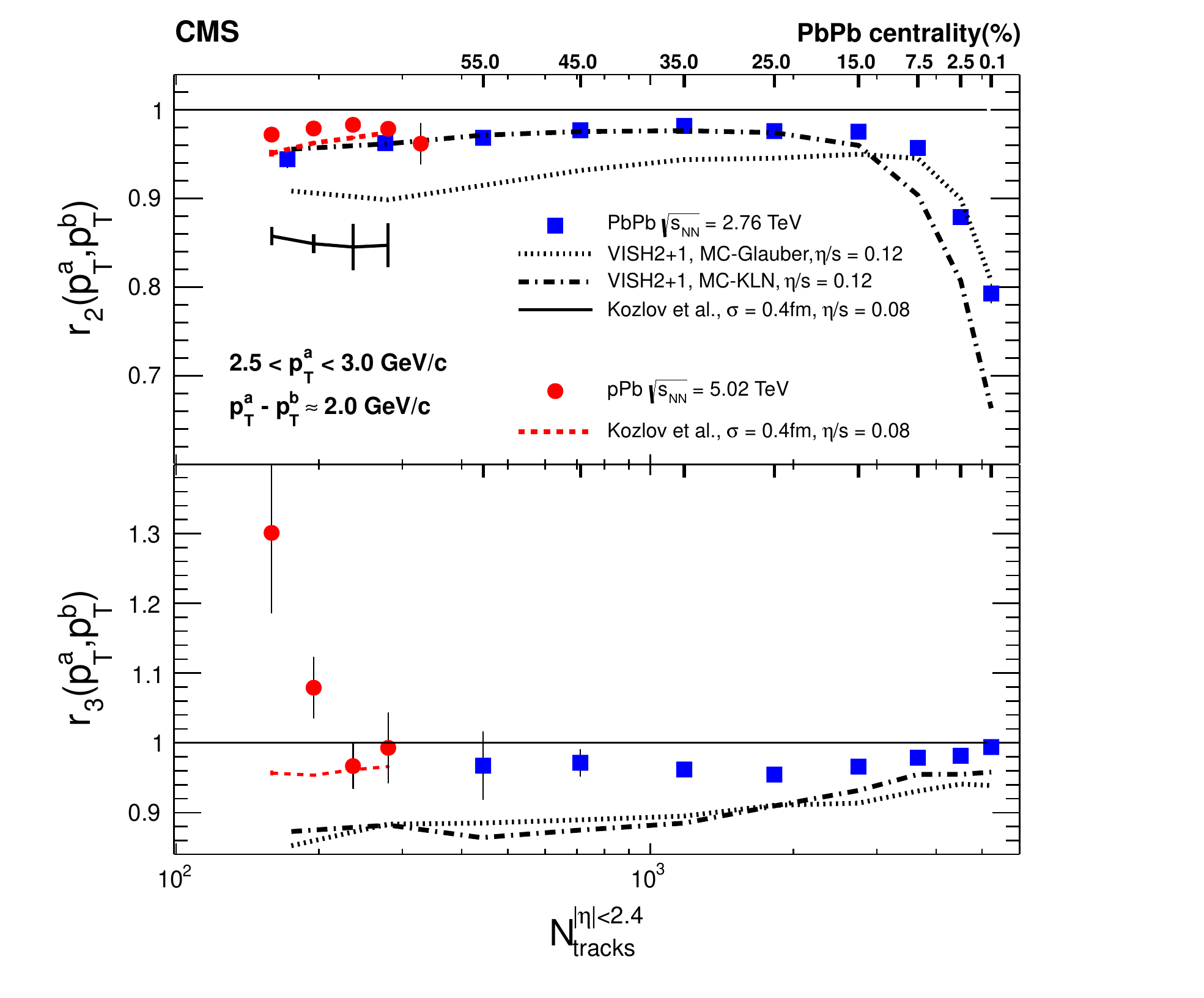}
    \caption{(Color online) The \pt-dependent factorization ratios, $r_2$
and $r_3$, as
    a function of event multiplicity in \pPb and \PbPb collisions.
    The curves show the calculations for \PbPb collisions from viscous
    hydrodynamics in Ref.~\cite{Heinz:2013bua}
    with MC-Glauber and MC-KLN initial condition models and $\eta/s=0.12$, and
    also hydrodynamic predictions for \PbPb and \pPb data
    in Ref.~\cite{Kozlov:2014fqa}.
    The horizontal solid lines denote the $r_{2}$ (top) and $r_{3}$ (bottom) value of unity.
    The error bars correspond to statistical uncertainties, while systematic
    uncertainties are negligible for the $r_n$ results and thus are not shown.
    }
    \label{fig:rn_cent}
\end{figure*}

Figure~\ref{fig:rn_pPb} shows a comparison between 5.02\TeV \pPb and 2.76\TeV peripheral \PbPb collisions over the same
multiplicity ranges. Because of the statistical limitation
of the \PbPb data, the multiplicity ranges used in Figs.~\ref{fig:r2_pPb} and \ref{fig:r3_pPb}
for \pPb data are combined into two \noff\ classes, $100 \leq \noff\ < 185$
(top) and $185 \leq \noff\ < 260$ (bottom).
At a similar \noff\ range, the magnitudes of factorization breakdown
in \pPb and \PbPb collisions depart from unity by less than 8\%, with
slightly smaller deviations for \pPb data, although
the statistical precision is limited. For both high-multiplicity \pPb and peripheral \PbPb collisions, the observed effect is significantly
smaller than that for
0--0.2\% centrality ultra-central \PbPb collisions (up to 20\% away from unity).
The similar behavior (e.g., \pt-dependence) of factorization data in \pPb to that in \PbPb collisions
may provide new insight into the possible
hydrodynamic flow
origin of long-range two-particle correlations in the \pPb system, particularly in providing
new information on the nature of initial-state fluctuations in a much smaller volume.

To study directly the multiplicity dependence of the effect
in \PbPb and \pPb collisions, the $r_2$ and $r_3$ results for $2.5<\pttrg<3.0$\GeVc
and $0.3<\ptass<0.5$\GeVc (where the difference between \pttrg and \ptass
is the greatest, $\pttrg-\ptass \approx 2$\GeVc) are shown in
Fig.~\ref{fig:rn_cent}, as a function
of event multiplicity in \pPb and \PbPb collisions.
Here, the number of tracks is still counted with $\abs{\eta}<2.4$ and $\pt>0.4$\GeVc but
corrected for the detector inefficiency, since a different track
reconstruction algorithm
is used for the \pPb and central \PbPb data. Additionally, at the top of the
figure, a centrality axis is shown which is applicable only
to PbPb collisions. The breakdown of factorization for $r_2$ in \PbPb events increases dramatically as the collisions become more central than
0--5\%,
while the effect in $r_3$ remains at the 2--3\% level, largely independent of centrality.
For more peripheral \PbPb events from 20\% to 80\% centrality, the deviation of $r_2$ from unity increases slightly from about 2 to 5\%.
Calculations using a hydrodynamic model in \PbPb collisions~\cite{Heinz:2013bua}
with MC-Glauber and MC-KLN initial conditions and $\eta/s = 0.12$ are also
shown as dotted and dash-dotted curves, respectively, as a
function of centrality. As pointed out earlier, neither of the two calculations can describe
the data quantitatively over the entire centrality range, although the qualitative
trend is reproduced. The $r_2$ values for \pPb show little multiplicity dependence,
consistent with hydrodynamic predictions in Ref.~\cite{Kozlov:2014fqa}.
The $r_{3}$ values for \pPb go significantly above unity at lower multiplicities,
because of the onset of nonflow correlations.
The discrepancy in the hydrodynamic calculations of $r_2$ for peripheral \PbPb collisions between Ref.~\cite{Heinz:2013bua} and Ref.~\cite{Kozlov:2014fqa} may
be related to differences in some model parameters (e.g., transverse size of the nucleon).
This should be investigated in the future.

Although the factorization results presented in this paper suggest a breakdown
of the assumption commonly applied in studying collective flow using two-particle
correlations (Eq.~(\ref{eq:fact})), previous $v_n$ measurements from the two-particle
method still remain valid. However, they should be more precisely interpreted as the $v_n$ values
obtained with respect to an averaged event plane using particles from a given kinematic
regime (usually over a wide \pt range). The studies in this paper also point out the
importance of applying the same conditions for theoretical calculations when comparing
to the experimental data.

\section{Pseudorapidity dependence of factorization breakdown}
\label{sec:eta}

\subsection{Analysis technique}
\label{sec:etaana}

In principle, the $\eta$-dependent factorization breakdown
and event plane angle fluctuations can be examined using a similar
formalism to Eq.~(\ref{r_n_def}) by replacing \pttrg and \ptass
with \etatrg and \etaass. However, the main issue with this approach
is that the requirement of $\abs{\deta}>2$ for removing short-range
two-particle correlations cannot be fulfilled anymore as the denominator
of the factorization ratio takes the $V_{n\Delta}(\etatrg,\etaass)$ components, where $\etatrg \approx \etaass$.
The correlation signal from collective flow is strongly contaminated
by short-range jet-like correlations. To avoid this problem, an
alternative observable is developed for the study of $\eta$-dependent
factorization, by taking advantage of the wide $\eta$ coverage of
the CMS tracker and HF calorimeters.

\begin{figure*}[t!h]
\centering
    \includegraphics[width=\cmsFigWidth]{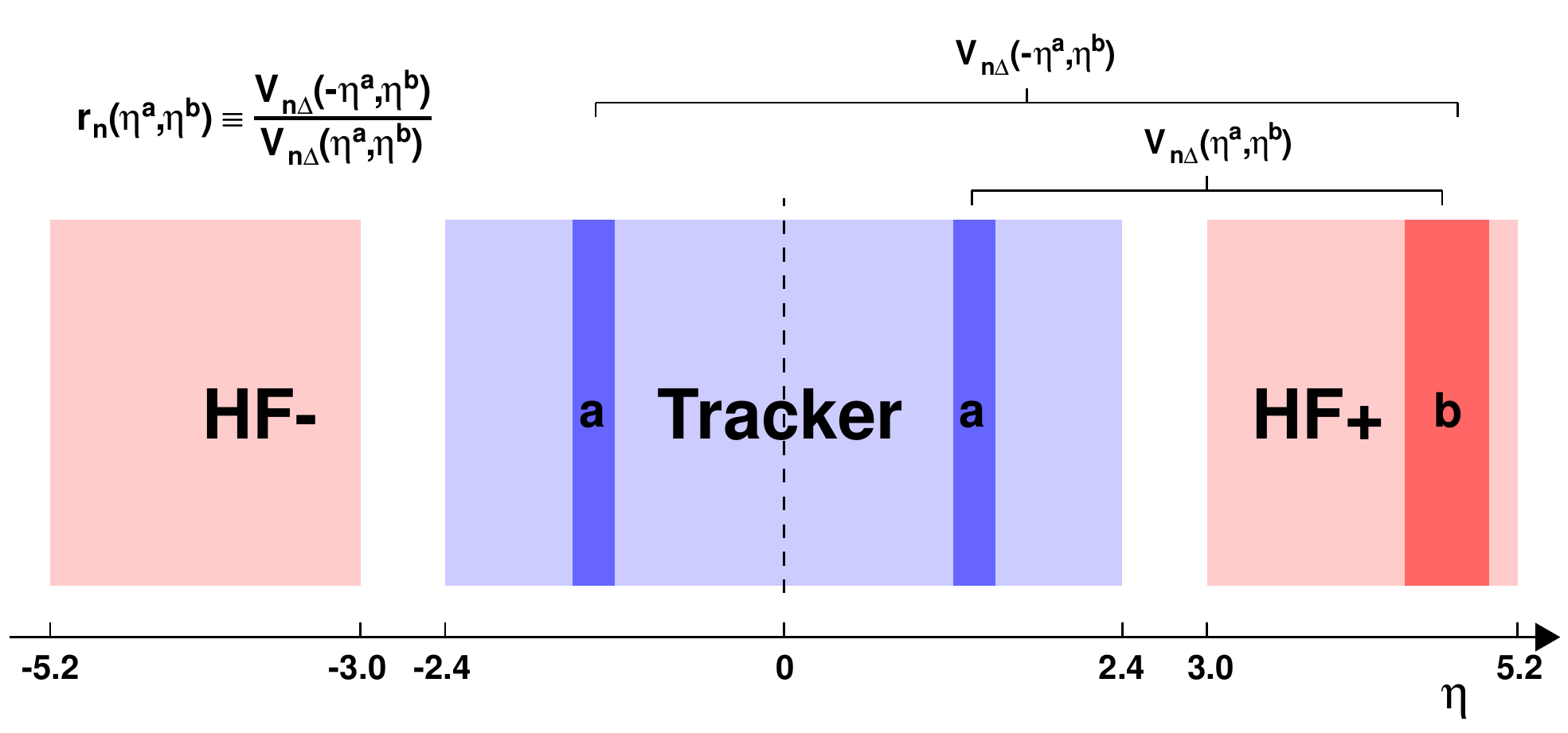}
    \caption{(Color online) A schematic illustrating the acceptance
coverage of the CMS tracker and HF
      calorimeters, and the procedure for deriving the $\eta$-dependent factorization ratio,
      $r_{n}(\etatrg,\etaass)$.
    }
    \label{fig:schematic}
\end{figure*}

The $\eta$-dependent factorization
ratio, $r_{n}(\etatrg,\etaass)$, is defined as
\begin{linenomath}
\begin{equation}
    \label{rn_eta}
    r_{n}(\etatrg,\etaass) \equiv \frac{V_{n\Delta}(-\etatrg,\etaass)}{V_{n\Delta}(\etatrg,\etaass)},
\end{equation}
\end{linenomath}
where $V_{n\Delta}(\etatrg,\etaass)$ is calculated in the same way
as Eq.~(\ref{average}) but for pairs of particles taken from varied
\etatrg and
\etaass regions in fixed \pttrg and \ptass ranges. Here, particle $a$ is chosen from charged
tracks with $0.3<\pttrg<3.0$\GeVc and $\abs{\etatrg}<2.4$, while particle $b$ is
selected from the HF calorimeter towers with the energy exceeding 1\GeV (with a total coverage of
$2.9<\abs{\eta}<5.2$) without any explicit transverse energy (\et) threshold for each tower.
With this approach, the $\eta$ values of both particles from a pair
can be varied over a wide range, while it is possible to ensure a large
$\eta$ gap by combining detector components covering central and forward $\eta$
regions. As illustrated by a schematic in Fig.~\ref{fig:schematic}, for $4.4<\etaass<5.0$
from the HF calorimeters, a minimum $\eta$ gap of 2 units
between a calorimeter tower and any charged particle from the
silicon tracker is guaranteed. Away-side back-to-back jet correlations could still
be present but they are shown to have a negligible contribution at low
\pt because of very
high multiplicities~\cite{Chatrchyan:2013nka}, especially in central \PbPb collisions. To account for any
occupancy effect of the HF detectors due to large granularities in $\eta$ and $\phi$,
each tower is weighted by its \ET value when calculating the average in
Eq.~(\ref{average}). For consistency, each track is also weighted by its \pt value.
The finite azimuthal resolution of the HF towers (0.349\unit{radians}) has
negligible effects to the $V_{n\Delta}$ calculation, which takes an \ET-weighted
average of 36 tower segments over a $2\pi$ coverage.

If, for each event, the event plane angle, $\Psi_n$, does vary for
particles produced at different $\eta$ regions, the following
relation for $r_{n}(\etatrg,\etaass)$ can be derived,
\begin{linenomath}
\begin{equation}
    \label{rn_eta1}
    r_{n}(\etatrg,\etaass) = \frac{\left\langle 
v_{n}(-\etatrg)v_{n}(\etaass)\cos\{n\big[\Psi_n(-\etatrg)-\Psi_n(\etaass)\big]\}\right\rangle}{\left\langle 
v_{n}(\etatrg)v_{n}(\etaass)\cos\{n\big[\Psi_n(\etatrg)-\Psi_n(\etaass)\big]\}\right\rangle}.
\end{equation}
\end{linenomath}
In symmetric collision systems like \PbPb, $v_n$ harmonics from symmetric
positive ($v_n({\etatrg})$) and negative ($v_n({-\etatrg})$) $\eta$
regions are identical after averaging over all events.
Therefore, Eq.~(\ref{rn_eta1}) can be approximated by
\begin{linenomath}
\begin{equation}
    \label{rn_eta2}
    r_{n}(\etatrg,\etaass) \approx \frac{\left\langle \cos\left[n\left(\Psi_n(-\etatrg)-\Psi_n(\etaass)\right)\right]\right\rangle}{\left\langle\cos\left[n\left(\Psi_n(\etatrg)-\Psi_n(\etaass)\right)\right]\right\rangle}.
\end{equation}
\end{linenomath}
Here, the approximation is due to the fact that the flow
magnitude ($v_n$) and the orientation angle ($\Psi_n$)
are inside the same averaging over all the events in the numerator of Eq.~(\ref{rn_eta1}).
As a result, $r_{n}(\etatrg,\etaass)$ represents a measurement of relative
event plane angle fluctuations in $\eta$ for planes separated by $\abs{\etatrg+\etaass}$
and $\abs{\etatrg-\etaass}$. Similar to $r_{n}(\pttrg,\ptass)$,
$r_{n}(\etatrg,\etaass)$ is equal to unity if the factorization holds but
factorization
breaks down in general in the presence of event plane fluctuations in $\eta$.

For an asymmetric collision system like \pPb, $v_n({\etatrg})$ and
$v_n({-\etatrg})$ are
not identical in general, and thus $\eta$-dependent event plane fluctuation effect
cannot be isolated in Eq.~(\ref{rn_eta1}). However, by taking a product of
$r_{n}(\etatrg,\etaass)$
and $r_{n}(-\etatrg,-\etaass)$, the $v_n$ terms can be removed,
\ifthenelse{\boolean{cms@external}}{
\begin{multline}
    \label{rn_eta2_pPb}
    \sqrt{r_{n}(\etatrg,\etaass) r_{n}(-\etatrg,-\etaass)} \approx\\
\sqrt{\frac{\left\langle
\cos\left[n\left(\Psi_n(-\etatrg)-\Psi_n(\etaass)\right)\right]\right\rangle}{\left\langle\cos\left[n\left(\Psi_n(\etatrg)
-\Psi_n(\etaass)\right)\right]\right\rangle}
\frac{\left\langle \cos\left[n\left(\Psi_n(\etatrg)-\Psi_n(-\etaass)\right)\right]\right\rangle}{\left\langle\cos\left[n\left(\Psi_n(-\etatrg)-\Psi_n(-\etaass)\right)\right]\right\rangle}}.
\end{multline}
}{
\begin{equation}
    \label{rn_eta2_pPb}
    \sqrt{r_{n}(\etatrg,\etaass) r_{n}(-\etatrg,-\etaass)} \approx
\sqrt{\frac{\left\langle
\cos\left[n\left(\Psi_n(-\etatrg)-\Psi_n(\etaass)\right)\right]\right\rangle}{\left\langle\cos\left[n\left(\Psi_n(\etatrg)
-\Psi_n(\etaass)\right)\right]\right\rangle}
\frac{\left\langle \cos\left[n\left(\Psi_n(\etatrg)-\Psi_n(-\etaass)\right)\right]\right\rangle}{\left\langle\cos\left[n\left(\Psi_n(-\etatrg)-\Psi_n(-\etaass)\right)\right]\right\rangle}}.
\end{equation}
}
In this way, the $\eta$-dependent event plane
angle fluctuations in \pPb collisions can also be studied.

\subsection{Results for \PbPb data}

\begin{figure*}[thb]
\centering
    \includegraphics[width=\linewidth]{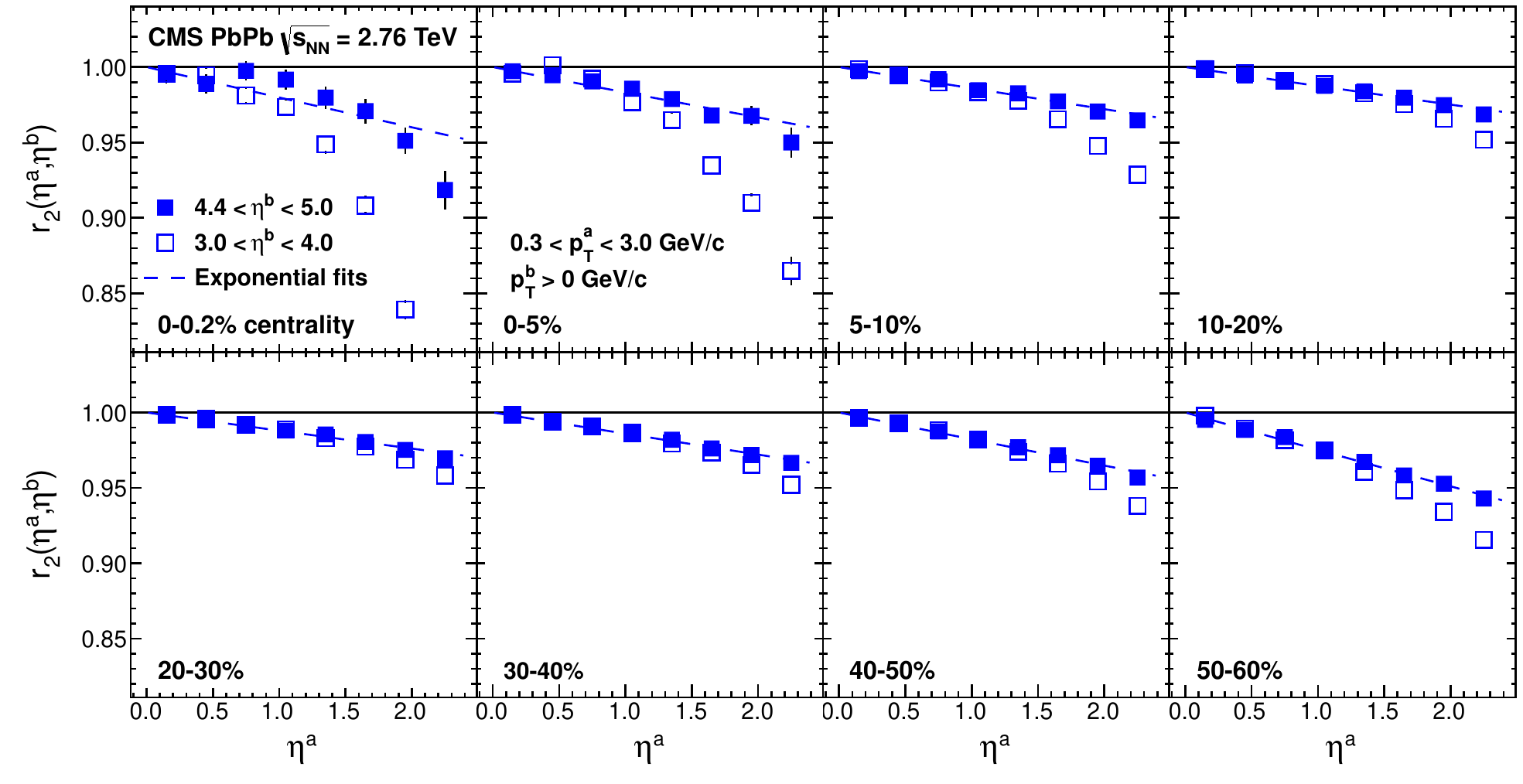}
    \caption{(Color online) The $\eta$-dependent factorization ratio,
$r_2$, as a function of \etatrg for $3.0<\etaass<4.0$
    and $4.4<\etaass<5.0$, averaged over $0.3<\pttrg<3.0$\GeVc,
    in eight centrality classes of \PbPb collisions at
$\sqrt{s_{NN}} = 2.76$\TeV.
    The curves correspond to fits to the data for $4.4<\etaass<5.0$ given by
Eq.~(\ref{epdeco_rn}). The horizontal solid lines denote the $r_{2}$ value of unity.
    The error bars correspond to statistical uncertainties, while systematic
    uncertainties are negligible for the $r_n$ results, and thus are not shown.
    }
    \label{fig:epetadeco_HI_r2}
\end{figure*}

\begin{figure*}[thb]
\centering
    \includegraphics[width=\linewidth]{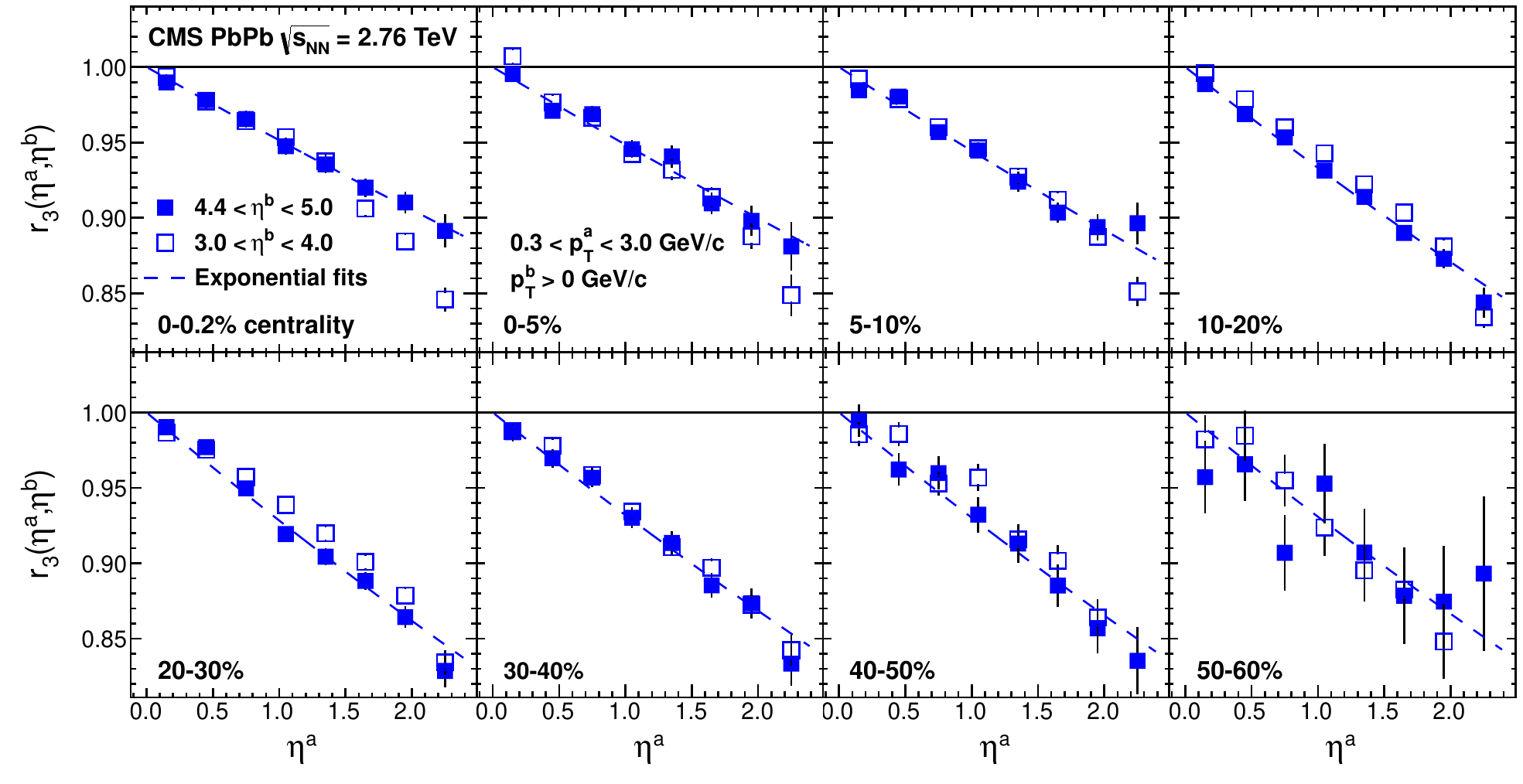}
    \caption{(Color online) Similar distributions as shown in
Fig.~\ref{fig:epetadeco_HI_r2},
but for the factorization ratio $r_3$.
    }
    \label{fig:epetadeco_HI_r3}
\end{figure*}

\begin{figure*}[thb]
\centering
    \includegraphics[width=\cmsFigWidth]{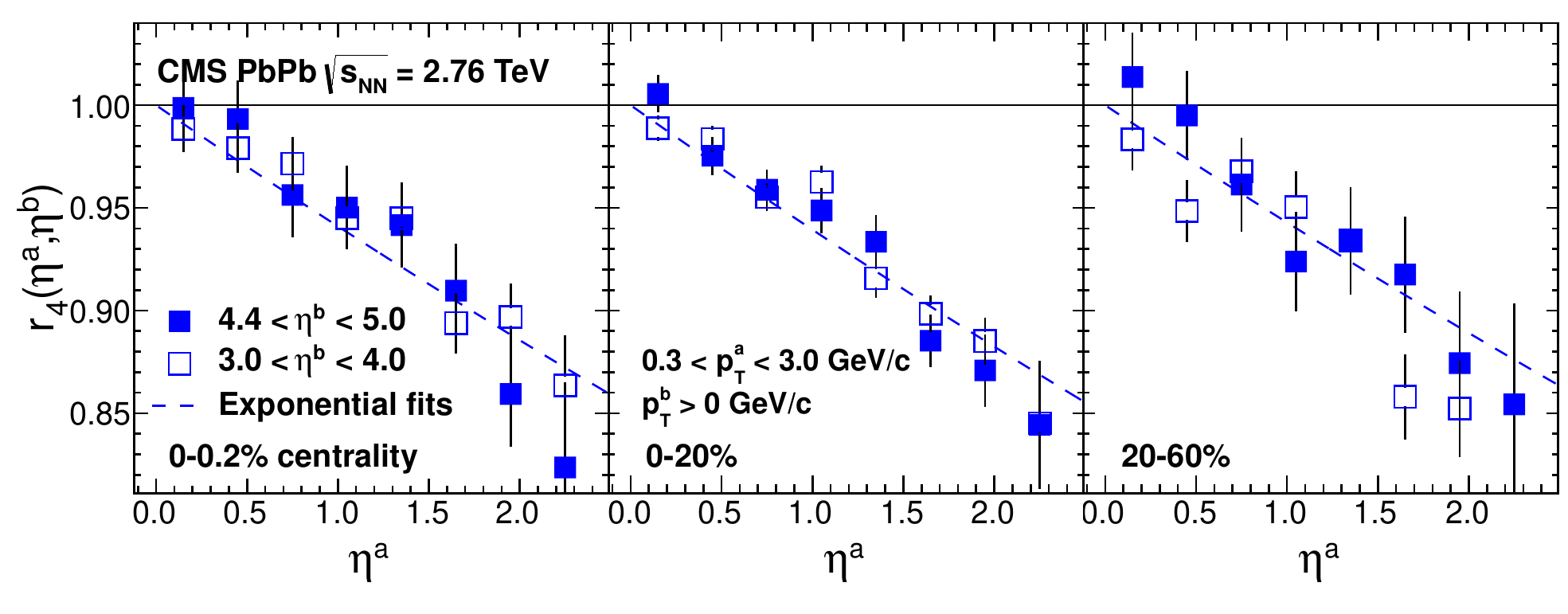}
    \caption{(Color online) Similar distributions as shown in
Fig.~\ref{fig:epetadeco_HI_r2},
but for the factorization ratio $r_4$
    in fewer centrality ranges.
    }
    \label{fig:epetadeco_HI_r4}
\end{figure*}

The results of $\eta$-dependent factorization ratios, $r_2$, $r_3$ and $r_4$
in \PbPb collisions at $\rootsNN = 2.76$\TeV
are shown in Figs.~\ref{fig:epetadeco_HI_r2}--\ref{fig:epetadeco_HI_r4}, as a function of \etatrg for
eight different centrality classes from 0--0.2\% to 50--60\% (except for $r_4$,
for which only three centrality classes are shown because of statistical limitations).
The $r_2(\etatrg,\etaass)$ values are calculated in \etatrg bins of 0.3 units, and the \etatrg value
at the center of each bin is used in the plots.
Data obtained with calorimeter tower $\eta$ ranges $3.0<\etaass<4.0$ and $4.4<\etaass<5.0$ are both presented.
Since \PbPb is a symmetric system, the $V_n(\etatrg,\etaass)$
and $V_n(-\etatrg,-\etaass)$ coefficients are combined before calculating the $r_n$
ratios in order to achieve the optimal statistical precision. Charged tracks within
$0.3<\pt<3.0$\GeVc and all calorimeter towers ($E > 1$\GeV) are used.
When \etatrg = 0, the $r_n$ value is equal to unity by construction since both the numerator
and denominator of $r_n$ have the same $\eta$ gap between particle $a$ and
$b$, as indicated in Eq.~(\ref{rn_eta2}).
As \etatrg increases, a significant decrease of $r_n$ below unity is observed,
which may suggest the presence of $\eta$-dependent event plane
angle fluctuations.

The $r_2$ values for $4.4<\etaass<5.0$ are found to decrease with \etatrg approximately
linearly for most of the centrality classes up to a few percent deviation
below unity
at $\etatrg \sim 2.4$. This behavior is slightly different for the most central 0--0.2\% events,
where the decrease of $r_2$ becomes more significant at $\etatrg \sim 1$.
For $3.0<\etaass<4.0$, the $r_2$ value exhibits a much stronger factorization breakdown effect
for an $\etatrg > 1$. This can be understood as the effect of
short-range jet-like correlations when the $\eta$ gap between two particles
is less than 2,
which increases the denominator of Eq.~(\ref{rn_eta}). However, for
$\etatrg<1$, the $r_2$ results
are found to be consistent with each other, independent of \etaass\ (except for 0--0.2\% centrality).
This demonstrates that contributions of short-range jet-like correlations are almost completely suppressed
if the requirement of $\abs{\Delta\eta}>2$ to both numerator and denominator of
$r_n(\etatrg,\etaass)$ is imposed.

The effect of $\eta$-dependent factorization breakdown is much stronger for
higher-order harmonics, $r_3$ and $r_4$, shown in Figs.~\ref{fig:epetadeco_HI_r3}
and \ref{fig:epetadeco_HI_r4}. For $r_3$, this trend is opposite to what is
observed for the \pt-dependent factorization ratio. For all centrality ranges
(including 0--0.2\%), an approximate linear dependence of $r_3$ and $r_4$ is seen.
Results from the two different \etaass\ ranges agree over most of the \etatrg range
within statistical uncertainties. This might suggest that short-range jet-like
correlations have much smaller effects on higher-order harmonics.

As observed in Figs.~\ref{fig:epetadeco_HI_r2}--\ref{fig:epetadeco_HI_r4},
the $r_n(\etatrg,\etaass)$ values are independent of \etaass, for \etatrg ranges where contributions of only long-range ($|\Delta\eta|>2$) correlations are included.
To quantify the dependence of $r_n$ values on \etatrg, a simple empirical parameterization is introduced:
\begin{equation}
    \label{epdeco_expo}
    \cos\left[n\left(\Psi_n(\etatrg)-\Psi_n(\etaass)\right)\right] =
\re^{-F^{\eta}_{n}\abs{\etatrg-\etaass}},
\end{equation}
which is based on the assumption that relative fluctuations between two event
plane angles depend only on their pseudorapidity difference. At small
$\Delta\eta$
values, the exponential function form can be approximated by a linear function
 of \deta, consistent with the observation in the data.
By plugging Eq.~(\ref{epdeco_expo}) into Eq.~(\ref{rn_eta2}), the $r_n$ can be expressed as
\begin{equation}
    \label{epdeco_rn}
    r_n(\etatrg,\etaass) \approx \re^{-2F^{\eta}_{n}\etatrg},
\end{equation}
which is independent of \etaass, consistent with the results in
Figs.~\ref{fig:epetadeco_HI_r2}--\ref{fig:epetadeco_HI_r4}. According to Eqs.~(\ref{epdeco_expo}) and (\ref{epdeco_rn}),
the $r_n(\etatrg,\etaass)$ also corresponds to a measurement of event plane fluctuations
between $\Psi_n(\etatrg)$ and $\Psi_n(-\etatrg)$,
\begin{equation}
    \label{rn_eta3}
    r_{n}(\etatrg,\etaass) \approx \left\langle \cos\left[n\left(\Psi_n(-\etatrg)-\Psi_n(\etatrg)\right)\right]\right\rangle.
\end{equation}
The $r_2$ data for $4.4<\etaass<5.0$ are well fit with a functional
form given by Eq.~(\ref{epdeco_rn}) for most centrality classes
($\chi^{2}/$(degree of freedom) $\sim$1), except for 0--0.2\% centrality, where
the $r_2$ value deviates from unity much faster as \etatrg increases.
Note that the parameter, $F^{\eta}_{n}$, is purely
empirical, without any clear physical meaning at present. It is introduced mainly for
quantitatively evaluating the centrality evolution of factorization
breakdown effect, as will be discussed later in Section~\ref{sec:comp_eta}.

\subsection{Results for \pPb data}

\begin{figure*}[thb]
\centering
    \includegraphics[width=\cmsFigWidth]{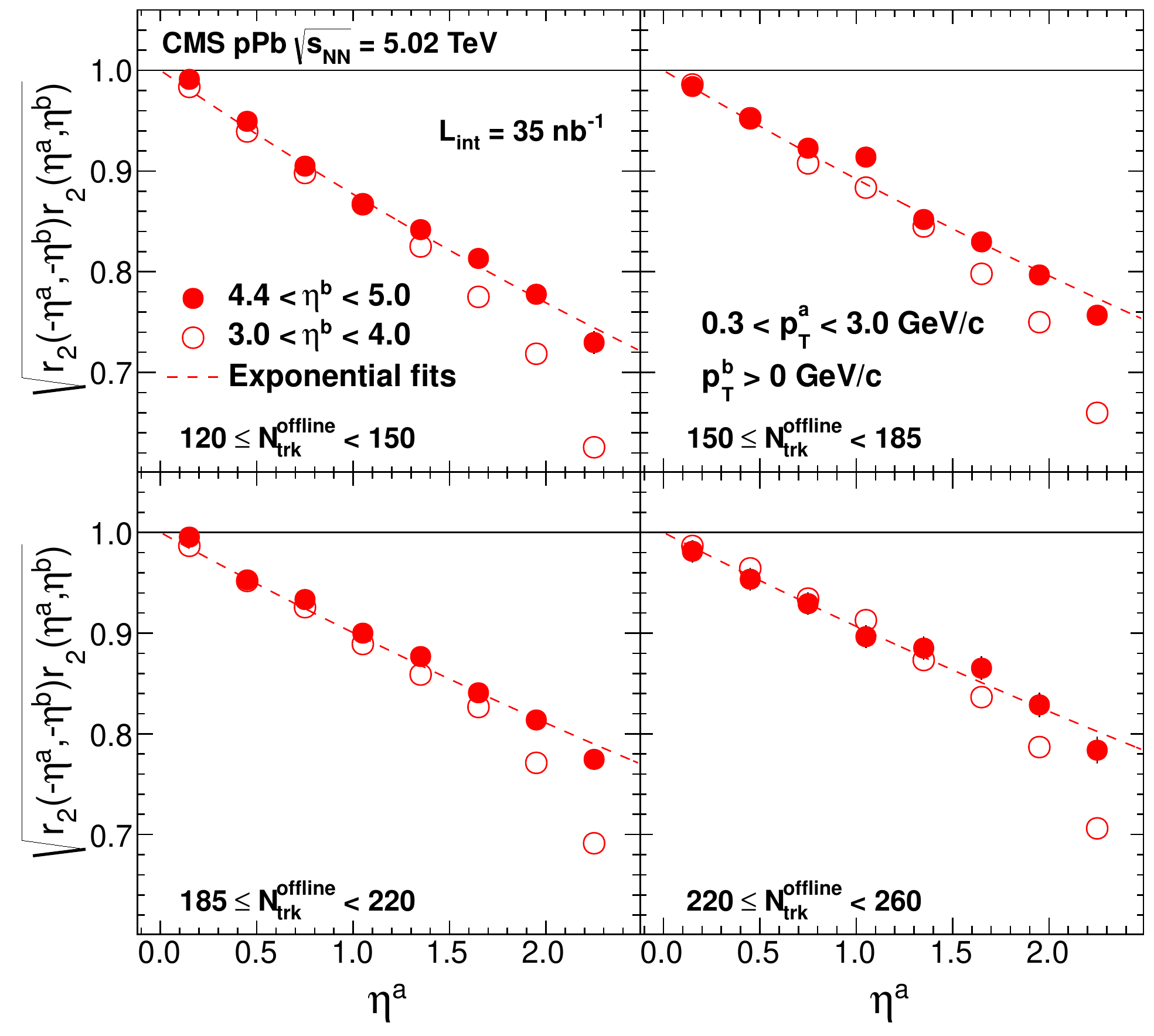}
    \caption{(Color online) The square root of the product of
factorization ratios,
    $\sqrt{r_{2}(\etatrg,\etaass) r_{2}(-\etatrg,-\etaass)}$,
    as a function of \etatrg for $3.0<\etaass<4.0$
    and $4.4<\etaass<5.0$, averaged over $0.3<\pttrg<3.0$\GeVc,
    in four multiplicity classes of \pPb collisions at $\rootsNN = 5.02\TeV$.
    The curves correspond to the fits to the data for $4.4<\etaass<5.0$
using Eq.~(\ref{epdeco_rn}). The horizontal solid lines denote the $r_{2}$ value of unity.
    The error bars correspond to statistical uncertainties, while systematic
    uncertainties are negligible for the $r_2$ results, and thus are not shown.
    }
    \label{fig:epetadeco_pPb_r2tot}
\end{figure*}

Studies of $\eta$-dependent factorization breakdown of two-particle correlations
are also performed in \pPb collisions at $\rootsNN = 5.02\TeV$ for four high-multiplicity
ranges, shown in Fig.~\ref{fig:epetadeco_pPb_r2tot} for the second-order
harmonics. Results for higher-order harmonics in \pPb cannot be obtained due to
statistical limitation.
As pointed out in Section \ref{sec:etaana},
because of asymmetry of pPb collisions in $\eta$, the factorization ratio, $r_{n}(\etatrg,\etaass)$,
is sensitive to asymmetry in the magnitude of $v_n$, and thus does not only
reflect the effect of event plane angle fluctuations. Therefore, the results are
presented as the square root of the product of $r_{n}(\etatrg,\etaass)$
and $r_{n}(-\etatrg,-\etaass)$, which is designed to remove the sensitivity to
the magnitude of $v_n$ (see Eq.~(\ref{rn_eta2_pPb}) for details). Similar
to
those in \PbPb collisions, two different $\eta$ ranges of HF towers, $3.0<\etaass<4.0$
and $4.4<\etaass<5.0$, are compared.

A significant breakdown of factorization in $\eta$ is also observed in \pPb collisions as \etatrg increases. Similar to the \PbPb results, the
factorization breakdown is approximately independent of \etaass\ for $\etatrg < 1$
for all multiplicity ranges but shows a much larger deviation from unity for $3.0<\etaass<4.0$
as \etatrg increases beyond one unit because of short-range correlations. The fits to the data for $4.4<\etaass<5.0$
using Eq.~(\ref{epdeco_rn}) are also shown; the data are well-described
over the
accessible $\etatrg$ range. It should be noted that the assumption made in
Eq.~(\ref{epdeco_expo})
is purely an empirical parameterization for quantifying the behavior of the data. Since \pPb collisions are asymmetric,
this assumption could be invalid. More detailed investigations on how $r_n$
depends on \etatrg and \etaass\ in the proton- and lead-going directions, respectively,
are needed in future work.

\subsection{Comparison of \pPb and \PbPb data}
\label{sec:comp_eta}

\begin{figure*}[thb]
\centering
    \includegraphics[width=\cmsFigWidth]{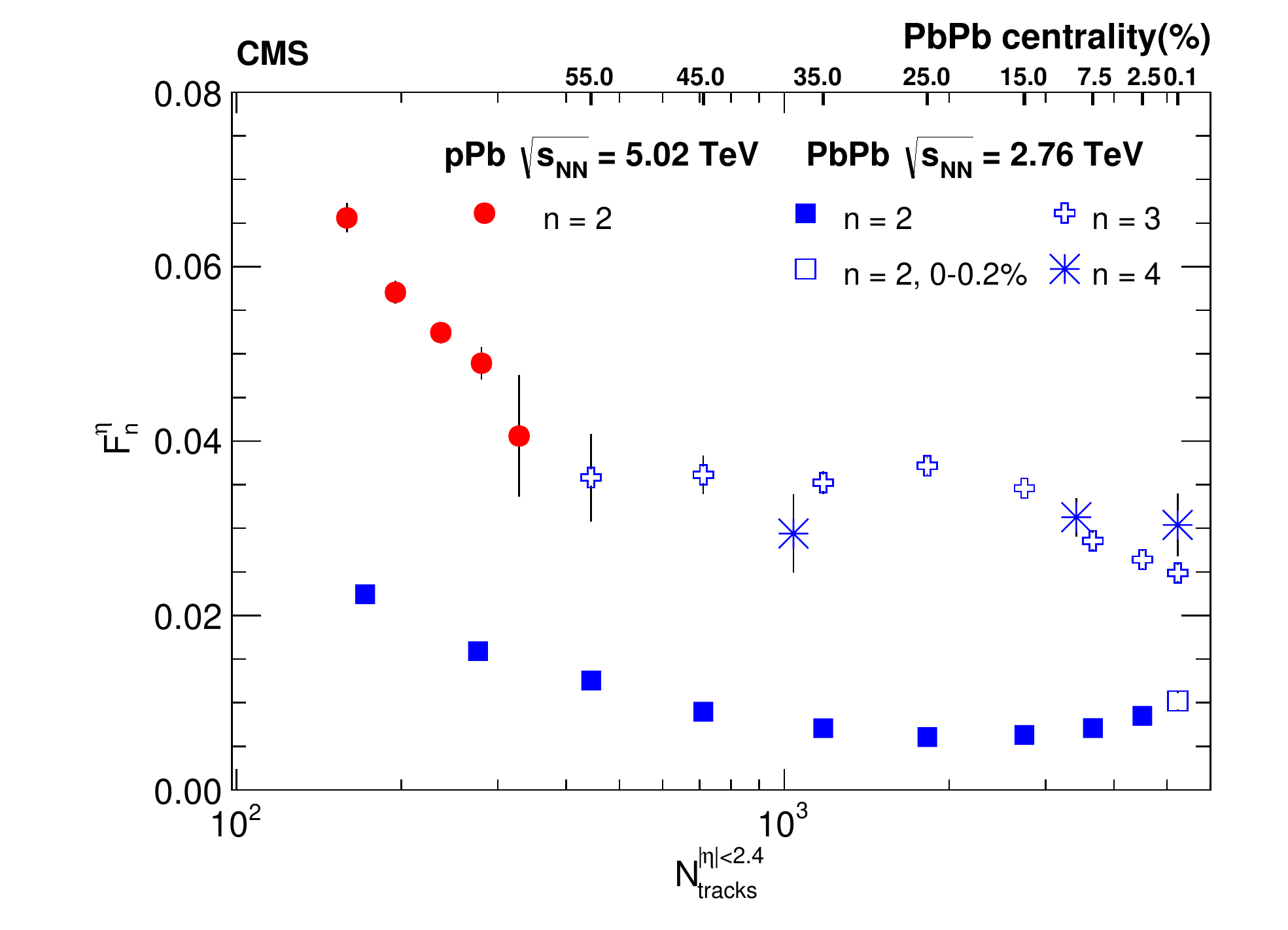}
    \caption{(Color online) The $F^{\eta}_{n}$ parameter as defined in
Eq.~(\ref{epdeco_rn})
              as a function of event multiplicity in \PbPb collisions at $\rootsNN = 2.76$\TeV for $n=2$--$4$
              and \pPb collisions at $\rootsNN = 5.02\TeV$ for $n=2$.
              The error bars correspond to statistical uncertainties, while systematic
              uncertainties are negligible for the $r_n$ results, and thus
are not shown.
    }
    \label{fig:epetadeco_HI_C}
\end{figure*}

The extracted $F^{\eta}_{n}$
parameters are plotted as a function of event multiplicity in Fig.~\ref{fig:epetadeco_HI_C},
in \pPb collisions for $n=2$ and \PbPb collisions for $n=2$--$4$.
The $F^{\eta}_{2}$
value reaches its minimum around midcentral ($\sim$20\%) \PbPb events, and increases significantly
for more peripheral \PbPb events and also for \pPb events, where the
relative fluctuations of $v_2$ are larger~\cite{Chatrchyan:2013kba}. Toward the most central
\PbPb events, the $F^{\eta}_{2}$ value also shows a tendency to increase slightly, although the $r_n$ data for 0--0.2\% centrality
are not well described by Eq.~(\ref{epdeco_rn}). At a similar multiplicity,
magnitudes of the $F^{\eta}_{2}$ parameter in \pPb are significantly larger than those in \PbPb,
and decrease with increasing event multiplicity.
In \PbPb collisions, a much stronger $\eta$-dependent factorization breakdown is
seen for higher-order harmonics than for the second order, as shown by the
$F^{\eta}_{3}$ and $F^{\eta}_{4}$ parameters.
There is little centrality dependence for $n=3$, except for the most central 0--20\% \PbPb collisions.
Within current statistical uncertainties, no centrality dependence is observed for $n=4$.

\section{Summary}

Factorization of azimuthal two-particle correlations into single-particle anisotropies
has been studied as a function of transverse momentum and pseudorapidity of each particle from a pair,
in \PbPb collisions at $\rootsNN = 2.76$\TeV and \pPb collisions at
$\rootsNN = 5.02\TeV$, over a wide multiplicity range.
The factorization assumption is found to be broken as a function of both \pt and $\eta$.
The effect of \pt\-dependent factorization breakdown for the second-order
Fourier harmonic is found to increase with the difference in \pt between the two
particles. The factorization breakdown reaches 20\% for the most central \PbPb collisions, while it decreases
rapidly for more peripheral collisions. The effect is significantly smaller (2--3\%) in
high-multiplicity \pPb collisions. In both \PbPb and \pPb samples over the full
centrality or multiplicity range, little effect is observed for the third order harmonic.
For the $\eta$ dependence, the observed
factorization breakdown shows an approximately linear increase with
the $\eta$ gap between two particles for all centrality and multiplicity
classes in \PbPb and \pPb collisions. The effect is weakest for mid-central
\PbPb\ events, but becomes larger for more central or peripheral \PbPb collisions,
and also for very high-multiplicity
\pPb\ collisions. Moreover, a much stronger $\eta$-dependent effect is seen
for the third- and fourth-order harmonics than the second-order harmonics
in \PbPb collisions. This relation between the second and third order
is opposite to that seen in the \pt-dependent factorization studies.
The observed factorization breakdown presented here does not invalidate
previous $v_n$ measurements. Instead, the previous values should be reinterpreted
as measuring anisotropies with respect to the event plane averaged over
a given kinematic region. Furthermore, it is important to compare data and
theoretical calculations following exactly the same procedure.

The factorization data have been compared to hydrodynamic calculations with fluctuating initial-state conditions.
The \pt-dependent factorization data are qualitatively described by
viscous hydrodynamic models, which are shown to be largely insensitive to
the value of shear viscosity to entropy density ratio of the medium.
This observation offers great promise
for using the factorization data to disentangle contributions of
the initial-state conditions and the medium's transport properties to
the observed collective flow phenomena in the final state.
The new studies of $\eta$-dependent factorization breakdown give an indication of
initial-state fluctuations along the longitudinal direction. This
will provide new insights into the longitudinal dynamics
of relativistic heavy ion collisions, and help improve the three-dimensional
modeling of the evolution of the strongly-coupled quark gluon medium.

\begin{acknowledgments}
We congratulate our colleagues in the CERN accelerator departments for the excellent performance of the LHC and thank the technical and administrative staffs at CERN and at other CMS institutes for their contributions to the success of the CMS effort. In addition, we gratefully acknowledge the computing centers and personnel of the Worldwide LHC Computing Grid for delivering so effectively the computing infrastructure essential to our analyses. Finally, we acknowledge the enduring support for the construction and operation of the LHC and the CMS detector provided by the following funding agencies: BMWFW and FWF (Austria); FNRS and FWO (Belgium); CNPq, CAPES, FAPERJ, and FAPESP (Brazil); MES (Bulgaria); CERN; CAS, MoST, and NSFC (China); COLCIENCIAS (Colombia); MSES and CSF (Croatia); RPF (Cyprus); MoER, ERC IUT and ERDF (Estonia); Academy of Finland, MEC, and HIP (Finland); CEA and CNRS/IN2P3 (France); BMBF, DFG, and HGF (Germany); GSRT (Greece); OTKA and NIH (Hungary); DAE and DST (India); IPM (Iran); SFI (Ireland); INFN (Italy); MSIP and NRF (Republic of Korea); LAS (Lithuania); MOE and UM (Malaysia); CINVESTAV, CONACYT, SEP, and UASLP-FAI (Mexico); MBIE (New Zealand); PAEC (Pakistan); MSHE and NSC (Poland); FCT (Portugal); JINR (Dubna); MON, RosAtom, RAS and RFBR (Russia); MESTD (Serbia); SEIDI and CPAN (Spain); Swiss Funding Agencies (Switzerland); MST (Taipei); ThEPCenter, IPST, STAR and NSTDA (Thailand); TUBITAK and TAEK (Turkey); NASU and SFFR (Ukraine); STFC (United Kingdom); DOE and NSF (USA).

Individuals have received support from the Marie-Curie program and the European Research Council and EPLANET (European Union); the Leventis Foundation; the A. P. Sloan Foundation; the Alexander von Humboldt Foundation; the Belgian Federal Science Policy Office; the Fonds pour la Formation \`a la Recherche dans l'Industrie et dans l'Agriculture (FRIA-Belgium); the Agentschap voor Innovatie door Wetenschap en Technologie (IWT-Belgium); the Ministry of Education, Youth and Sports (MEYS) of the Czech Republic; the Council of Science and Industrial Research, India; the HOMING PLUS program of the Foundation for Polish Science, cofinanced from European Union, Regional Development Fund; the Compagnia di San Paolo (Torino); the Consorzio per la Fisica (Trieste); MIUR project 20108T4XTM (Italy); the Thalis and Aristeia programs cofinanced by EU-ESF and the Greek NSRF; and the National Priorities Research Program by Qatar National Research Fund.
\end{acknowledgments}

\bibliography{auto_generated}

\providecommand{\href}[2]{#2}\begingroup\raggedright\begin{thebibliography}{10}%
\makeatletter
\providecommand{\hrefCMSnoop }[0]{\@secondoftwo}%
\makeatother
\providecommand{\doi}{\texttt{doi:}\begingroup \urlstyle{tt}\Url}

\bibitem{BRAHMS}
\hrefCMSnoop {}{{BRAHMS} Collaboration, ``{Quark gluon plasma and color glass
  condensate at RHIC? The perspective from the BRAHMS experiment}'',} \textit{
  Nucl. Phys. A} \textbf{ 757} (2005) 1,
  \href{http://dx.doi.org/10.1016/j.nuclphysa.2005.02.130}{\doi{10.1016/j.nuclphysa.2005.02.130}},
\href{http://www.arXiv.org/abs/nucl-ex/0410020}{\texttt{arXiv:nucl-ex/0410020}}.

\bibitem{PHOBOS}
\hrefCMSnoop {}{{PHOBOS} Collaboration, ``{The PHOBOS perspective on
  discoveries at RHIC}'',} \textit{ Nucl. Phys. A} \textbf{ 757} (2005) 28,
  \href{http://dx.doi.org/10.1016/j.nuclphysa.2005.03.084}{\doi{10.1016/j.nuclphysa.2005.03.084}},
\href{http://www.arXiv.org/abs/nucl-ex/0410022}{\texttt{arXiv:nucl-ex/0410022}}.

\bibitem{STAR}
\hrefCMSnoop {}{{STAR} Collaboration, ``{Experimental and theoretical
  challenges in the search for the quark gluon plasma: The STAR Collaboration's
  critical assessment of the evidence from RHIC collisions}'',} \textit{ Nucl.
  Phys. A} \textbf{ 757} (2005) 102,
  \href{http://dx.doi.org/10.1016/j.nuclphysa.2005.03.085}{\doi{10.1016/j.nuclphysa.2005.03.085}},
\href{http://www.arXiv.org/abs/nucl-ex/0501009}{\texttt{arXiv:nucl-ex/0501009}}.

\bibitem{PHENIX}
\hrefCMSnoop {}{{PHENIX} Collaboration, ``{Formation of dense partonic matter
  in relativistic nucleus-nucleus collisions at RHIC: Experimental evaluation
  by the PHENIX collaboration}'',} \textit{ Nucl. Phys. A} \textbf{ 757} (2005)
  184,
  \href{http://dx.doi.org/10.1016/j.nuclphysa.2005.03.086}{\doi{10.1016/j.nuclphysa.2005.03.086}},
\href{http://www.arXiv.org/abs/nucl-ex/0410003}{\texttt{arXiv:nucl-ex/0410003}}.

\bibitem{Aamodt:2011by}
\hrefCMSnoop {}{{ALICE Collaboration}, ``{Harmonic decomposition of
  two-particle angular correlations in Pb-Pb collisions at \rootsNN\ =
  2.76\TeV}'',} \textit{ Phys. Lett. B} \textbf{ 708} (2012) 249,
  \href{http://dx.doi.org/10.1016/j.physletb.2012.01.060}{\doi{10.1016/j.physletb.2012.01.060}},
\href{http://www.arXiv.org/abs/1109.2501}{\texttt{arXiv:1109.2501}}.

\bibitem{Abelev:2014pua}
\hrefCMSnoop {}{{ALICE Collaboration}, ``{Elliptic flow of identified hadrons
  in Pb-Pb collisions at \rootsNN\ = 2.76\TeV}'',} (2014).
\href{http://www.arXiv.org/abs/1405.4632}{\texttt{arXiv:1405.4632}}.

\bibitem{ATLAS:2012at}
\hrefCMSnoop {}{{ATLAS Collaboration}, ``{Measurement of the azimuthal
  anisotropy for charged particle production in \rootsNN\ = 2.76\TeV\ lead-lead
  collisions with the ATLAS detector}'',} \textit{ Phys. Rev. C} \textbf{ 86}
  (2012) 014907,
  \href{http://dx.doi.org/10.1103/PhysRevC.86.014907}{\doi{10.1103/PhysRevC.86.014907}},
\href{http://www.arXiv.org/abs/1203.3087}{\texttt{arXiv:1203.3087}}.

\bibitem{Aad:2013xma}
\hrefCMSnoop {}{{ATLAS Collaboration}, ``{Measurement of the distributions of
  event-by-event flow harmonics in lead-lead collisions at \rootsNN\ = 2.76\TeV
  with the ATLAS detector at the LHC}'',} \textit{ JHEP} \textbf{ 11} (2013)
  183,
  \href{http://dx.doi.org/10.1007/JHEP11(2013)183}{\doi{10.1007/JHEP11(2013)183}},
\href{http://www.arXiv.org/abs/1305.2942}{\texttt{arXiv:1305.2942}}.

\bibitem{Aad:2014fla}
\hrefCMSnoop {}{{ATLAS Collaboration}, ``{Measurement of event-plane
  correlations in \rootsNN\ = 2.76\TeV\ lead-lead collisions with the ATLAS
  detector}'',} \textit{ Phys. Rev. C} \textbf{ 90} (2014) 024905,
  \href{http://dx.doi.org/10.1103/PhysRevC.90.024905}{\doi{10.1103/PhysRevC.90.024905}},
\href{http://www.arXiv.org/abs/1403.0489}{\texttt{arXiv:1403.0489}}.

\bibitem{Chatrchyan:2012wg}
\hrefCMSnoop {}{{CMS Collaboration}, ``{Centrality dependence of dihadron
  correlations and azimuthal anisotropy harmonics in PbPb collisions at
  \rootsNN\ = 2.76 TeV}'',} \textit{ Eur. Phys. J. C} \textbf{ 72} (2012) 2012,
  \href{http://dx.doi.org/10.1140/epjc/s10052-012-2012-3}{\doi{10.1140/epjc/s10052-012-2012-3}},
\href{http://www.arXiv.org/abs/1201.3158}{\texttt{arXiv:1201.3158}}.

\bibitem{Chatrchyan:2012ta}
\hrefCMSnoop {}{{CMS Collaboration}, ``{Measurement of the elliptic anisotropy
  of charged particles produced in PbPb collisions at nucleon-nucleon
  center-of-mass energy = 2.76 TeV}'',} \textit{ Phys. Rev. C} \textbf{ 87}
  (2013) 014902,
  \href{http://dx.doi.org/10.1103/PhysRevC.87.014902}{\doi{10.1103/PhysRevC.87.014902}},
\href{http://www.arXiv.org/abs/1204.1409}{\texttt{arXiv:1204.1409}}.

\bibitem{Chatrchyan:2013kba}
\hrefCMSnoop {}{{CMS Collaboration}, ``{Measurement of higher-order harmonic
  azimuthal anisotropy in PbPb collisions at \rootsNN\ = 2.76\TeV}'',} \textit{
  Phys. Rev. C} \textbf{ 89} (2014) 044906,
  \href{http://dx.doi.org/10.1103/PhysRevC.89.044906}{\doi{10.1103/PhysRevC.89.044906}},
\href{http://www.arXiv.org/abs/1310.8651}{\texttt{arXiv:1310.8651}}.

\bibitem{CMS:2013bza}
\hrefCMSnoop {}{{CMS Collaboration}, ``{Studies of azimuthal dihadron
  correlations in ultra-central PbPb collisions at \rootsNN\ = 2.76 TeV}'',}
  \textit{ JHEP} \textbf{ 02} (2014) 088,
  \href{http://dx.doi.org/10.1007/JHEP02(2014)088}{\doi{10.1007/JHEP02(2014)088}},
\href{http://www.arXiv.org/abs/1312.1845}{\texttt{arXiv:1312.1845}}.

\bibitem{Ollitrault:1993ba}
\hrefCMSnoop {}{J.-Y. Ollitrault, ``{Determination of the reaction plane in
  ultrarelativistic nuclear collisions}'',} \textit{ Phys. Rev. D} \textbf{ 48}
  (1993) 1132,
  \href{http://dx.doi.org/10.1103/PhysRevD.48.1132}{\doi{10.1103/PhysRevD.48.1132}},
\href{http://www.arXiv.org/abs/hep-ph/9303247}{\texttt{arXiv:hep-ph/9303247}}.

\bibitem{Voloshin:1994mz}
\hrefCMSnoop {}{S.~Voloshin and Y.~Zhang, ``{Flow study in relativistic nuclear
  collisions by Fourier expansion of azimuthal particle distributions}'',}
  \textit{ Z. Phys. C} \textbf{ 70} (1996) 665,
  \href{http://dx.doi.org/10.1007/s002880050141}{\doi{10.1007/s002880050141}},
\href{http://www.arXiv.org/abs/hep-ph/9407282}{\texttt{arXiv:hep-ph/9407282}}.

\bibitem{Poskanzer:1998yz}
\hrefCMSnoop {}{A.~M. Poskanzer and S.~A. Voloshin, ``{Methods for analyzing
  anisotropic flow in relativistic nuclear collisions}'',} \textit{ Phys. Rev.
  C} \textbf{ 58} (1998) 1671,
  \href{http://dx.doi.org/10.1103/PhysRevC.58.1671}{\doi{10.1103/PhysRevC.58.1671}},
\href{http://www.arXiv.org/abs/nucl-ex/9805001}{\texttt{arXiv:nucl-ex/9805001}}.

\bibitem{Alver:2010gr}
\hrefCMSnoop {}{B.~Alver and G.~Roland, ``{Collision geometry fluctuations and
  triangular flow in heavy-ion collisions}'',} \textit{ Phys. Rev. C} \textbf{
  81} (2010) 054905,
  \href{http://dx.doi.org/10.1103/PhysRevC.81.054905}{\doi{10.1103/PhysRevC.81.054905}},
  \href{http://www.arXiv.org/abs/1003.0194}{\texttt{arXiv:1003.0194}}.
[Erratum: \DOI{10.1103/PhysRevC.82.039903}].

\bibitem{Alver:2010dn}
\hrefCMSnoop {}{B.~H. Alver, C.~Gombeaud, M.~Luzum, and J.-Y. Ollitrault,
  ``{Triangular flow in hydrodynamics and transport theory}'',} \textit{ Phys.
  Rev. C} \textbf{ 82} (2010) 034913,
  \href{http://dx.doi.org/10.1103/PhysRevC.82.034913}{\doi{10.1103/PhysRevC.82.034913}},
\href{http://www.arXiv.org/abs/1007.5469}{\texttt{arXiv:1007.5469}}.

\bibitem{Schenke:2010rr}
\hrefCMSnoop {}{B.~Schenke, S.~Jeon, and C.~Gale, ``Elliptic and triangular
  flow in event-by-event {D=3+1} viscous hydrodynamics'',} \textit{ Phys. Rev.
  Lett.} \textbf{ 106} (2011) 042301,
  \href{http://dx.doi.org/10.1103/PhysRevLett.106.042301}{\doi{10.1103/PhysRevLett.106.042301}},
\href{http://www.arXiv.org/abs/1009.3244}{\texttt{arXiv:1009.3244}}.

\bibitem{Qiu:2011hf}
\hrefCMSnoop {}{Z.~Qiu, C.~Shen, and U.~Heinz, ``{Hydrodynamic elliptic and
  triangular flow in Pb-Pb collisions at \rootsNN\ = 2.76\TeV}'',} \textit{
  Phys. Lett. B} \textbf{ 707} (2012) 151,
  \href{http://dx.doi.org/10.1016/j.physletb.2011.12.041}{\doi{10.1016/j.physletb.2011.12.041}},
\href{http://www.arXiv.org/abs/1110.3033}{\texttt{arXiv:1110.3033}}.

\bibitem{Danielewicz:1985hn}
\hrefCMSnoop {}{P.~Danielewicz and G.~Odyniec, ``Transverse momentum analysis
  of collective motion in relativistic nuclear collisions'',} \textit{ Phys.
  Lett. B} \textbf{ 157} (1985) 146,
\href{http://dx.doi.org/10.1016/0370-2693(85)91535-7}{\doi{10.1016/0370-2693(85)91535-7}}.

\bibitem{Chatrchyan:2013nka}
\hrefCMSnoop {}{{CMS Collaboration}, ``{Multiplicity and transverse momentum
  dependence of two- and four-particle correlations in pPb and PbPb
  collisions}'',} \textit{ Phys. Lett. B} \textbf{ 724} (2013) 213,
  \href{http://dx.doi.org/10.1016/j.physletb.2013.06.028}{\doi{10.1016/j.physletb.2013.06.028}},
\href{http://www.arXiv.org/abs/1305.0609}{\texttt{arXiv:1305.0609}}.

\bibitem{Gardim:2012im}
\hrefCMSnoop {}{F.~G. Gardim, F.~Grassi, M.~Luzum, and J.-Y. Ollitrault,
  ``{Breaking of factorization of two-particle correlations in
  hydrodynamics}'',} \textit{ Phys. Rev. C} \textbf{ 87} (2013) 031901,
  \href{http://dx.doi.org/10.1103/PhysRevC.87.031901}{\doi{10.1103/PhysRevC.87.031901}},
\href{http://www.arXiv.org/abs/1211.0989}{\texttt{arXiv:1211.0989}}.

\bibitem{Heinz:2013bua}
\hrefCMSnoop {}{U.~Heinz, Z.~Qiu, and C.~Shen, ``{Fluctuating flow angles and
  anisotropic flow measurements}'',} \textit{ Phys. Rev. C} \textbf{ 87} (2013)
  034913,
  \href{http://dx.doi.org/10.1103/PhysRevC.87.034913}{\doi{10.1103/PhysRevC.87.034913}},
\href{http://www.arXiv.org/abs/1302.3535}{\texttt{arXiv:1302.3535}}.

\bibitem{Kozlov:2014fqa}
I.~Kozlov\hrefCMSnoop {}{ {et~al.}, ``{Transverse momentum structure of pair
  correlations as a signature of collective behavior in small collision
  systems}'',} (2014).
\href{http://www.arXiv.org/abs/1405.3976}{\texttt{arXiv:1405.3976}}.

\bibitem{Floerchinger:2013rya}
\hrefCMSnoop {}{S.~Floerchinger and U.~A. Wiedemann, ``{Mode-by-mode fluid
  dynamics for relativistic heavy ion collisions}'',} \textit{ Phys. Lett. B}
  \textbf{ 728} (2014) 407,
  \href{http://dx.doi.org/10.1016/j.physletb.2013.12.025}{\doi{10.1016/j.physletb.2013.12.025}},
\href{http://www.arXiv.org/abs/1307.3453}{\texttt{arXiv:1307.3453}}.

\bibitem{ColemanSmith:2012ka}
\hrefCMSnoop {}{C.~E. Coleman-Smith, H.~Petersen, and R.~L. Wolpert,
  ``{Classification of initial state granularity via 2d Fourier expansion}'',}
  \textit{ J. Phys. G} \textbf{ 40} (2013) 095103,
  \href{http://dx.doi.org/10.1088/0954-3899/40/9/095103}{\doi{10.1088/0954-3899/40/9/095103}},
\href{http://www.arXiv.org/abs/1204.5774}{\texttt{arXiv:1204.5774}}.

\bibitem{Khachatryan:2010gv}
\hrefCMSnoop {}{{CMS Collaboration}, ``Observation of long-range near-side
  angular correlations in proton-proton collisions at the {LHC}'',} \textit{
  JHEP} \textbf{ 09} (2010) 091,
  \href{http://dx.doi.org/10.1007/JHEP09(2010)091}{\doi{10.1007/JHEP09(2010)091}},
\href{http://www.arXiv.org/abs/1009.4122}{\texttt{arXiv:1009.4122}}.

\bibitem{CMS:2012qk}
\hrefCMSnoop {}{{CMS Collaboration}, ``{Observation of long-range near-side
  angular correlations in proton-lead collisions at the LHC}'',} \textit{ Phys.
  Lett. B} \textbf{ 718} (2013) 795,
  \href{http://dx.doi.org/10.1016/j.physletb.2012.11.025}{\doi{10.1016/j.physletb.2012.11.025}},
\href{http://www.arXiv.org/abs/1210.5482}{\texttt{arXiv:1210.5482}}.

\bibitem{alice:2012qe}
\hrefCMSnoop {}{{ALICE Collaboration}, ``{Long-range angular correlations on
  the near and away side in \pPb\ collisions at \rootsNN\ = 5.02\TeV }'',}
  \textit{ Phys. Lett. B} \textbf{ 719} (2013) 29,
  \href{http://dx.doi.org/10.1016/j.physletb.2013.01.012}{\doi{10.1016/j.physletb.2013.01.012}},
\href{http://www.arXiv.org/abs/1212.2001}{\texttt{arXiv:1212.2001}}.

\bibitem{Aad:2012gla}
\hrefCMSnoop {}{{ATLAS Collaboration}, ``{Observation of Associated Near-Side
  and Away-Side Long-Range Correlations in \rootsNN\ = 5.02\TeV\ Proton-Lead
  Collisions with the ATLAS Detector}'',} \textit{ Phys. Rev. Lett.} \textbf{
  110} (2013) 182302,
  \href{http://dx.doi.org/10.1103/PhysRevLett.110.182302}{\doi{10.1103/PhysRevLett.110.182302}},
\href{http://www.arXiv.org/abs/1212.5198}{\texttt{arXiv:1212.5198}}.

\bibitem{Aad:2014lta}
\hrefCMSnoop {}{{ATLAS Collaboration}, ``{Measurement of long-range
  pseudorapidity correlations and azimuthal harmonics in \rootsNN\ = 5.02\TeV\
  proton-lead collisions with the ATLAS detector}'',} \textit{ Phys. Rev. C}
  \textbf{ 90} (2014) 044906,
  \href{http://dx.doi.org/10.1103/PhysRevC.90.044906}{\doi{10.1103/PhysRevC.90.044906}},
\href{http://www.arXiv.org/abs/1409.1792}{\texttt{arXiv:1409.1792}}.

\bibitem{Bozek:2010vz}
\hrefCMSnoop {}{P.~Bozek, W.~Broniowski, and J.~Moreira, ``{Torqued fireballs
  in relativistic heavy-ion collisions}'',} \textit{ Phys. Rev. C} \textbf{ 83}
  (2011) 034911,
  \href{http://dx.doi.org/10.1103/PhysRevC.83.034911}{\doi{10.1103/PhysRevC.83.034911}},
\href{http://www.arXiv.org/abs/1011.3354}{\texttt{arXiv:1011.3354}}.

\bibitem{Pang:2014pxa}
L.-G. Pang\hrefCMSnoop {}{ {et~al.}, ``{Longitudinal Decorrelation of
  Anisotropic Flows in Heavy-ion Collisions at the LHC}'',} (2014).
\href{http://www.arXiv.org/abs/1410.8690}{\texttt{arXiv:1410.8690}}.

\bibitem{Xiao:2012uw}
\hrefCMSnoop {}{K.~Xiao, F.~Liu, and F.~Wang, ``{Event-plane decorrelation over
  pseudo-rapidity and its effect on azimuthal anisotropy measurement in
  relativistic heavy-ion collisions}'',} \textit{ Phys. Rev. C} \textbf{ 87}
  (2013) 011901,
  \href{http://dx.doi.org/10.1103/PhysRevC.87.011901}{\doi{10.1103/PhysRevC.87.011901}},
\href{http://www.arXiv.org/abs/1208.1195}{\texttt{arXiv:1208.1195}}.

\bibitem{Jia:2014vja}
\hrefCMSnoop {}{J.~Jia and P.~Huo, ``{A method for studying the rapidity
  fluctuation and decorrelation of harmonic flow in heavy-ion collisions}'',}
  \textit{ Phys. Rev. C} \textbf{ 90} (2014) 034905,
  \href{http://dx.doi.org/10.1103/PhysRevC.90.034905}{\doi{10.1103/PhysRevC.90.034905}},
\href{http://www.arXiv.org/abs/1402.6680}{\texttt{arXiv:1402.6680}}.

\bibitem{JINST}
\hrefCMSnoop {}{{CMS Collaboration}, ``{The CMS experiment at the CERN LHC}'',}
  \textit{ JINST} \textbf{ 3} (2008) S08004,
\href{http://dx.doi.org/10.1088/1748-0221/3/08/S08004}{\doi{10.1088/1748-0221/3/08/S08004}}.

\bibitem{GEANT4}
\hrefCMSnoop {}{{GEANT4} Collaboration, ``{GEANT4}---a simulation toolkit'',}
  \textit{ Nucl. Instrum. Meth. A} \textbf{ 506} (2003) 250,
\href{http://dx.doi.org/10.1016/S0168-9002(03)01368-8}{\doi{10.1016/S0168-9002(03)01368-8}}.

\bibitem{Djuvsland:2010qs}
\hrefCMSnoop {}{{\O}.~Djuvsland and J.~Nystrand, ``{Single and double
  photonuclear excitations in Pb+Pb collisions at \rootsNN\ = 2.76 TeV at the
  CERN Large Hadron Collider}'',} \textit{ Phys. Rev. C} \textbf{ 83} (2011)
  041901,
  \href{http://dx.doi.org/10.1103/PhysRevC.83.041901}{\doi{10.1103/PhysRevC.83.041901}},
\href{http://www.arXiv.org/abs/1011.4908}{\texttt{arXiv:1011.4908}}.

\bibitem{Lokhtin:2005px}
\hrefCMSnoop {}{I.~P. Lokhtin and A.~M. Snigirev, ``{A model of jet quenching
  in ultrarelativistic heavy ion collisions and high-\pt\ hadron spectra at
  RHIC}'',} \textit{ Eur. Phys. J. C} \textbf{ 45} (2006) 211,
  \href{http://dx.doi.org/10.1140/epjc/s2005-02426-3}{\doi{10.1140/epjc/s2005-02426-3}},
\href{http://www.arXiv.org/abs/hep-ph/0506189}{\texttt{arXiv:hep-ph/0506189}}.

\bibitem{Porteboeuf:2010um}
\hrefCMSnoop {}{S.~Porteboeuf, T.~Pierog, and K.~Werner, ``{Producing Hard
  Processes Regarding the Complete Event: The EPOS Event Generator}'',} (2010).
\href{http://www.arXiv.org/abs/1006.2967}{\texttt{arXiv:1006.2967}}.

\bibitem{Gyulassy:1994ew}
\hrefCMSnoop {}{M.~Gyulassy and X.-N. Wang, ``{HIJING 1.0: A Monte Carlo
  program for parton and particle production in high-energy hadronic and
  nuclear collisions}'',} \textit{ Comput. Phys. Commun.} \textbf{ 83} (1994)
  307,
  \href{http://dx.doi.org/10.1016/0010-4655(94)90057-4}{\doi{10.1016/0010-4655(94)90057-4}},
\href{http://www.arXiv.org/abs/nucl-th/9502021}{\texttt{arXiv:nucl-th/9502021}}.

\bibitem{Chatrchyan:2014fea}
\hrefCMSnoop {}{{CMS Collaboration}, ``{Description and performance of track
  and primary-vertex reconstruction with the CMS tracker}'',} \textit{ JINST}
  \textbf{ 9} (2014)
  \href{http://dx.doi.org/10.1088/1748-0221/9/10/P10009}{\doi{10.1088/1748-0221/9/10/P10009}},
\href{http://www.arXiv.org/abs/1405.6569}{\texttt{arXiv:1405.6569}}.

\bibitem{Chatrchyan:2011eka}
\hrefCMSnoop {}{{CMS Collaboration}, ``{Long-range and short-range dihadron
  angular correlations in central PbPb collisions at a nucleon-nucleon center
  of mass energy of 2.76 TeV}'',} \textit{ JHEP} \textbf{ 07} (2011) 076,
  \href{http://dx.doi.org/10.1007/JHEP07(2011)076}{\doi{10.1007/JHEP07(2011)076}},
\href{http://www.arXiv.org/abs/1105.2438}{\texttt{arXiv:1105.2438}}.

\bibitem{Khachatryan:2014jra}
\hrefCMSnoop {}{{CMS Collaboration}, ``{Long-range two-particle correlations of
  strange hadrons with charged particles in pPb and PbPb collisions at LHC
  energies}'',} \textit{ Phys. Lett. B} \textbf{ 742} (2015) 200,
  \href{http://dx.doi.org/10.1016/j.physletb.2015.01.034}{\doi{10.1016/j.physletb.2015.01.034}},
\href{http://www.arXiv.org/abs/1409.3392}{\texttt{arXiv:1409.3392}}.

\bibitem{glauber}
\hrefCMSnoop {}{M.~L. Miller, K.~Reygers, S.~J. Sanders, and P.~Steinberg,
  ``{Glauber modeling in high energy nuclear collisions}'',} \textit{ Ann. Rev.
  Nucl. Part. Sci.} \textbf{ 57} (2007) 205,
  \href{http://dx.doi.org/10.1146/annurev.nucl.57.090506.123020}{\doi{10.1146/annurev.nucl.57.090506.123020}},
\href{http://www.arXiv.org/abs/nucl-ex/0701025}{\texttt{arXiv:nucl-ex/0701025}}.

\bibitem{Alver:Glauber}
\hrefCMSnoop {}{B.~Alver, M.~Baker, C.~Loizides, and P.~Steinberg, ``{The
  PHOBOS Glauber Monte Carlo}'',} (2008).
  \href{http://www.arXiv.org/abs/0805.4411}{\texttt{arXiv:0805.4411}}.

\bibitem{Drescher:2006pi}
A.~Adil\hrefCMSnoop {}{ {et~al.}, ``{The eccentricity in heavy-ion collisions
  from color glass condensate initial conditions}'',} \textit{ Phys. Rev. C}
  \textbf{ 74} (2006) 044905,
  \href{http://dx.doi.org/10.1103/PhysRevC.74.044905}{\doi{10.1103/PhysRevC.74.044905}},
\href{http://www.arXiv.org/abs/nucl-th/0605012}{\texttt{arXiv:nucl-th/0605012}}.

\end{thebibliography}\endgroup

\cleardoublepage \appendix\section{The CMS Collaboration \label{app:collab}}\begin{sloppypar}\hyphenpenalty=5000\widowpenalty=500\clubpenalty=5000\textbf{Yerevan Physics Institute,  Yerevan,  Armenia}\\*[0pt]
V.~Khachatryan, A.M.~Sirunyan, A.~Tumasyan
\vskip\cmsinstskip
\textbf{Institut f\"{u}r Hochenergiephysik der OeAW,  Wien,  Austria}\\*[0pt]
W.~Adam, E.~Asilar, T.~Bergauer, J.~Brandstetter, M.~Dragicevic, J.~Er\"{o}, M.~Flechl, M.~Friedl, R.~Fr\"{u}hwirth\cmsAuthorMark{1}, V.M.~Ghete, C.~Hartl, N.~H\"{o}rmann, J.~Hrubec, M.~Jeitler\cmsAuthorMark{1}, W.~Kiesenhofer, V.~Kn\"{u}nz, A.~K\"{o}nig, M.~Krammer\cmsAuthorMark{1}, I.~Kr\"{a}tschmer, D.~Liko, I.~Mikulec, D.~Rabady\cmsAuthorMark{2}, B.~Rahbaran, H.~Rohringer, J.~Schieck\cmsAuthorMark{1}, R.~Sch\"{o}fbeck, J.~Strauss, W.~Treberer-Treberspurg, W.~Waltenberger, C.-E.~Wulz\cmsAuthorMark{1}
% ...
\vskip\cmsinstskip
% ...
1:~~Also at Vienna University of Technology, Vienna, Austria\\
2:~~Also at CERN, European Organization for Nuclear Research, Geneva, Switzerland\\
% ...

\end{sloppypar}
\end{document}